\documentclass[twocolumn,epjc3]{svjour3}
\usepackage{graphics}

\usepackage{cuted}
\usepackage{flushend}

\usepackage{microtype}
\usepackage{bbm}
\usepackage{amsmath}
\usepackage{mathtools}
\usepackage{amssymb}
\usepackage{bbold}
\usepackage{mathrsfs}       
\usepackage[pdftex]{graphicx}
\usepackage{pgfplots}
\pgfplotsset{width=10cm,compat=1.9}
\usepackage{xargs}       
\usepackage{wasysym}
\usepackage{engrec}
\usepackage{enumerate}
\usepackage{appendix}
\usepackage{empheq}
\usepackage{float}
\usepackage{stmaryrd}
\usepackage[parfill]{parskip}
\usepackage[pdftex,breaklinks]{hyperref}
\usepackage{titletoc}
\usepackage{makecell}
\usepackage{lscape}
\usepackage{algorithm}
\usepackage{algorithmicx}
\usepackage{tensor}
\usepackage{algpseudocode}
\usepackage{blkarray}
\usepackage{multirow}
\usepackage{bigstrut}
\usetikzlibrary{plotmarks}
\usepackage{braket}
\usepackage{cancel}
\usepackage{simplewick}
\usepackage[export]{adjustbox}

\usepackage{simpler-wick}
\usepackage{bm}

\allowdisplaybreaks

\begin{document}
\title{\textit{Ab initio} description of monopole resonances in light- and medium-mass nuclei}
\subtitle{III. Moments evaluation in \textit{ab initio} PGCM calculations}
\author{A. Porro\thanksref{ad:tud,ad:emmi,ad:saclay} 
\and T. Duguet\thanksref{ad:saclay,ad:kul}
\and J.-P. Ebran\thanksref{ad:dam,ad:dam_u}
\and M. Frosini\thanksref{ad:cadarache}
\and R. Roth\thanksref{ad:tud,ad:darm2}
\and V. Som\`a\thanksref{ad:saclay}}

\institute{
\label{ad:tud}
Technische Universit\"at Darmstadt, Department of Physics, 64289 Darmstadt, Germany
\and
\label{ad:emmi}
ExtreMe Matter Institute EMMI, GSI Helmholtzzentrum f\"ur Schwerionenforschung GmbH, 64291 Darmstadt, Germany
\and
\label{ad:saclay}
IRFU, CEA, Universit\'e Paris-Saclay, 91191 Gif-sur-Yvette, France 
\and
\label{ad:kul}
KU Leuven, Department of Physics and Astronomy, Instituut voor Kern- en Stralingsfysica, 3001 Leuven, Belgium
\and
\label{ad:dam}
CEA, DAM, DIF, 91297 Arpajon, France
\and
\label{ad:dam_u}
Universit\'e Paris-Saclay, CEA, Laboratoire Mati\`ere en Conditions Extr\^emes, 91680 Bruy\`eres-le-Ch\^atel, France
\and
\label{ad:cadarache}
CEA, DES, IRESNE, DER, SPRC, 13108 Saint-Paul-l\`es-Durance, France
\and
\label{ad:darm2}
Helmholtz Forschungsakademie Hessen f\"ur FAIR, GSI Helmholtzzentrum, 64289 Darmstadt, Germany
}

\date{Received: \today{} / Revised version: date}

\maketitle
%
%
\begin{abstract}
The paper is the third of a series dedicated to the ab initio description of monopole giant resonances in mid-mass closed- and open-shell nuclei via the so-called projected generator coordinate method. The present focus is on the computation of the moments $m_k$ of the monopole strength distribution, which are used to quantify its centroid energy and dispersion. First, the capacity to compute low-order moments via two different methods is developed and benchmarked for the $m_1$ moment. Second, the impact of the angular momentum projection on the centroid energy and dispersion of the monopole strength is analysed before comparing the results to those obtained from consistent quasi-particle random phase approximation calculations. Next, the so-called energy weighted sum rule (EWSR) is investigated. First, the appropriate ESWR in the center-of-mass frame is derived analytically. Second, the exhaustion of the intrinsic EWSR is tested in order to quantify the (unwanted) local-gauge symmetry breaking of the presently employed chiral effective field theory ($\chi$EFT) interactions. Finally, the infinite nuclear matter incompressibility associated with the employed $\chi$EFT interactions is extracted by extrapolating the finite-nucleus incompressibility computed from the monopole centroid energy.
\end{abstract}

\section{Introduction}

The study of giant resonances (GRs) provides valuable insights into the structural and dynamical properties of atomic nuclei. In particular, the characteristics of the isoscalar giant monopole resonance (ISGMR or GMR for brevity here) and of the isovector giant dipole resonance (IVGDR) not only deepen our comprehension of nuclear structure but also have implications for the modelisation of several astrophysical systems. This is the case, for instance, of the description of core-collapse supernov\ae~explosions and neutron stars mergers, both phenomena being associated to the nucleosynthesis of heavy elements and the behavior of nuclear matter under extreme conditions.

This article is the third (Paper III) of a series of four addressing the properties of the GMR in closed- and open-shell nuclei from an \textit{ab initio} standpoint using the so-called projected generator coordinate method (PGCM). While the first paper (Paper I)~\cite{Porro24a} detailed the uncertainty budget associated to PGCM calculations of monopole and quadrupole responses, the second paper (Paper II)~\cite{Porro:2024tzt} focused on the GMR properties of $^{16}$O, $^{24}$Mg, $^{28}$Si and $^{46}$Ti. Two-dimensional PGCM calculations were shown to account well for the fragmented monopole response of (rather) light doubly open-shell nuclei thanks to their capacity (i) to capture the impact of the intrinsic static quadrupole deformation and of its fluctuations on the position of the breathing mode (typically at play in spherical nuclei), (ii) to describe in a refined way the coupling between the GMR and the giant quadrupole resonance (GQR) mechanism responsible for the appearance of an additional component in the GMR of intrinsically-deformed nuclei and (iii) to seize anharmonic effects that were shown to be significant in light systems.

The present paper focuses on the computation of the moments $m_k$ of the monopole strength distribution in order to quantify its main characteristics such as its centroid energy and dispersion. Furthermore, the first moment $m_1$ leads to the so-called energy-weighted sum rule (EWSR) that is used to extract experimental strength functions. Also, the inverse-energy weighted sum rule (IEWSR) associated with the moment $m_{-1}$ delivers, when applied to the dipole response, the so-called dipole polarizability that is relevant to the computation of radiative capture cross sections. Finally, the centroid energy of the monopole strength distribution gives access to the nucleus-dependent nuclear compressibility $K_A$ that can eventually be linked to the nuclear matter incompressibility $K_{\infty}$. The latter quantity is a key characteristic of the nuclear equation of state (EOS) and, as such, has a clear interest for several astrophysical applications. 

The moments of a strength function can be computed in two ways. The first one involves an explicit sum over excited states and matrix elements of the simple one-body excitation operator $F$. 
The second one does involve the expectation value of a complicated many-body operator, but in the sole ground state. The first approach is presently denoted as the {\it sum over excited states} (SOES) method whereas the second one is referred to as the {\it ground-state expectation value} (GSEV) method. For a given many-body method, the agreement between the two approaches constitutes an internal-consistency test to pass\footnote{It is a necessary but not sufficient condition for the response function associated with the operator $F$ to be a good approximation of the exact one.} in order to correctly describe the excitation mode defined by the operator $F$. 

In this context, the formal capacity to compute low-order moments via the GSEV approach is developed in Sec.~\ref{sec:intro} and \ref{sec:mom_expl}. 
Based on such an advancement, and after briefly introducing the numerical setting in Sec.~\ref{technical}, the SOES and GSEV approaches to $m_1$ are compared in Sec.~\ref{sec:mom_meth_comp} using the PGCM monopole responses of $^{16}$O, $^{24}$Mg, $^{28}$Si and $^{46}$Ti.
Sec.~\ref{sec:mom_amp} further discusses the impact of angular-momentum projection in the SOES approach.
Next, PGCM moments are compared in Sec.~\ref{sec:mom_correlation} to those obtained via the quasi-particle random phase approximation (QRPA), the goal being to complement the study of Paper II dedicated to the monopole strength function by focusing on its global characteristics. 
Section~\ref{sec:ewsr_ex} focuses on the EWSR. First, it is demonstrated that the textbook expression of the EWSR must be corrected for the fact that nuclear excitations of interest are \textit{intrinsic} excitations in the center-of-mass frame. Second, the exhaustion of the intrinsic EWSR is tested in order to quantify the (unwanted) local-gauge symmetry breaking of the presently employed chiral Hamiltonian. 
Eventually, Sec.~\ref{sec:K} is dedicated to accessing $K_A$ in $^{16}$O, $^{24}$Mg, $^{28}$Si and $^{46}$Ti. The computed values are then employed to extract $K_{\infty}$ and verify if the result thus obtained is consistent with empirical expectations. This constitutes an important test for the chiral Hamiltonian under present use.  
The main findings of this work are summarised in Sec.~\ref{sec:concl} whereas a set of technical appendices complement the main body of the paper. 

\section{Moments of the strength function}
\label{sec:intro}

In many respects, the present section follows Ref.~\cite{Bohigas79a}. By convention, all operators at play are redefined in such a way that their expectation value in the ground state is subtracted, i.e. for a given operator $Q$ its rescaled companion is introduced as\footnote{Notice that Eq. \eqref{eq:convention} is no longer relevant as soon as one deals with commutators, since it is immediate to show that
$[A-\braket{A},B-\braket{B}]=[A,B]$.
}
\begin{equation}\label{eq:convention}
    \textbf{Q}\equiv Q-\braket{\Psi^{\sigma_0}_0|Q|\Psi^{\sigma_0}_0}\,.
\end{equation}

\subsection{Definition}

The strength function associated with a generic operator $F$ reads as
\begin{equation}
    S(E)\equiv\sum_{\nu \sigma}|\braket{\Psi^{\sigma}_\nu|\textbf{F}|\Psi^{\sigma_0}_0}|^2 \, \delta(E^{\sigma}_\nu-E^{\sigma_0}_0-E) \, , \label{eq:strength}
\end{equation}
where $| \Psi^{\sigma_0}_0 \rangle$ ($E^{\sigma_0}_0$) denotes the ground state (energy) whereas $\{| \Psi^{\sigma}_\nu \rangle (E^{\sigma}_\nu), \nu = 1, \ldots, \nu_{\text{max}} \}$ designates the set of excited states (energies). The superscript $\sigma\equiv(JM\Pi NZ)$ characterises the symmetry quantum numbers carried by the eigenstates, i.e. the angular momentum $J$ and its projection $M$, the parity $\Pi=\pm1$ as well as neutron $N$ and proton $Z$ numbers.

The \textit{k}-th moment of the strength distribution\footnote{From a mathematical standpoint, the moment $m_k$ constitutes the $k$-th moment of a discretized probability distribution associated with the transition generated by $F$. Moments of a physical strength function are not guaranteed to be finite. The fact that it is indeed the case or not depends on mathematical characteristics of the Hamiltonian, e.g. of inter-nucleon interactions~\cite{Bohigas79a}.} associated with the operator $\textbf{F}$ is defined as\footnote{Except for $m_0$, and as it will become evident below, there is in fact no difference in using $\textbf{F}$ or $F$ in Eq. \eqref{eq:def_k}.}
\begin{equation}
    m_k\equiv\int_{0}^{+\infty}E^kS(E)dE \, . \label{defmoments}
\end{equation}

\subsection{Mean energy(ies) and dispersion}

Two sets of quantities having the dimension of an energy are introduced according to
\begin{subequations}
\label{eq:ave_ene}
    \begin{align}
    	\label{eq:mom_av_def}
        \bar{E}_k&\equiv\frac{m_k}{m_{k-1}}\,,\\
        \tilde{E}_k&\equiv\bigg(\frac{m_k}{m_{k-2}}\bigg)^{1/2}\,.
    \end{align}
\end{subequations}
They coincide for all $k$'s if the strength distribution is concentrated in a single peak. The degree to which they differ reflects the fragmentation of the distribution. By definition, the average value of the energy distribution is given by
\begin{subequations}
\label{eq:ave_ene_new}
\begin{equation}
    \bar{E}_1=\frac{m_1}{m_0}\,.\label{eq:mid_en}
\end{equation}
In this work the following energy averages are also employed
\begin{align}
    \tilde{E}_1&=\sqrt{\frac{m_1}{m_{-1}}}\,,\label{eq:low_en}\\
    \tilde{E}_3&=\sqrt{\frac{m_3}{m_1}}\,.\;\label{eq:high_en}
\end{align}
\end{subequations}
Compared to the centroid energy $\bar{E}_1$, the scaled (constrained) energy $\tilde{E}_3$ ($\tilde{E}_1$)  is more sensitive to the high (low) energy part of the strength.

As shown in \ref{sec:schwartz}, the moments entertain the set of inequalities
\begin{equation}\label{eq:ineq}
    \dots\geq\frac{m_{k+2}}{m_{k+1}}\geq\sqrt{\frac{m_{k+2}}{m_k}}\geq\frac{m_{k+1}}{m_k}\geq\sqrt{\frac{m_{k+1}}{m_{k-1}}}\geq\dots\,,
\end{equation}
providing a practical tool to set boundaries on a specific moment in case it cannot be easily computed\footnote{From a practical standpoint, Eq.~\eqref{eq:ineq} holds if the involved moments are all computed within the same approximation scheme.}. Thanks to these inequalities, the variance of the strength distribution is shown to satisfy
\begin{equation}
    \sigma^2=\frac{m_2}{m_0}-\bigg(\frac{m_1}{m_0}\bigg)^2\geq0\,.
\end{equation}

\subsection{SOES formulation}

Inserting Eq.~\eqref{eq:strength} into Eq.~\eqref{defmoments} delivers the expression
\begin{align}
m_k &\equiv \sum_{\nu\sigma}(E^{\sigma}_\nu-E^{\sigma_0}_0)^k|\braket{\Psi^{\sigma}_\nu|\textbf{F}|\Psi^{\sigma_0}_0}|^2 \, ,\label{eq:def_k}
\end{align}
requiring the knowledge of excited states of the system. Equation~\eqref{eq:def_k} constitutes the SOES approach to the moments computation.  

\subsection{GSEV formulation}
\label{sec:comm_op}

By means of the identity resolution on the $A$-body Hilbert space $\mathcal{H}_A$
\begin{equation}
	\label{eq:id_res_mom}
\mathbb{1}=\sum_{\nu\sigma}\ket{\Psi^{\sigma}_\nu}\bra{\Psi^{\sigma}_\nu}\,,
\end{equation} 
Eq. \eqref{eq:def_k} can be rewritten as a ground-state expectation value
\begin{align}\label{eq:def_init}
m_k&=\sum_\nu(E^{\sigma}_\nu-E^{\sigma_0}_0)^k\braket{\Psi^{\sigma_0}_0|\textbf{F}|\Psi^{\sigma}_\nu}\braket{\Psi^{\sigma}_\nu|\textbf{F}|\Psi^{\sigma_0}_0}\nonumber\\
&=\sum_\nu\braket{\Psi^{\sigma_0}_0|\textbf{F}({H}-E^{\sigma_0}_0)^k|\Psi^{\sigma}_\nu}\braket{\Psi^{\sigma}_\nu|\textbf{F}|\Psi^{\sigma_0}_0}\nonumber\\
&=\braket{\Psi^{\sigma_0}_0|\textbf{F}({H}-E^{\sigma_0}_0)^k\textbf{F}|\Psi^{\sigma_0}_0}\,.
\end{align} 
Computing moments via Eq.~\eqref{eq:def_init} constitutes the GSEV method based on the expectation value of a complicated operator in the sole ground state. 

Clearly, the complexity of the operator at play in  Eq.~\eqref{eq:def_init} increases with $|k|$. For $k\geq 0$ the many-body rank increases with $k$ whereas for $k<0$ it further involves a non-trivial inversion.

\subsection{Moment operators}
\label{MomentOP}

Positive moments can be re-expressed in more convenient forms by invoking the appropriate definition of moment operators. As shown in \ref{sec:momentderivation} moments with $k\geq0$ can be further rewritten as
\begin{equation}\label{eq:mk_gen}
    m_k=(-1)^{i}\braket{\Psi^{\sigma_0}_0|C_iC_j|\Psi^{\sigma_0}_0}
\end{equation} 
with
\begin{align}\label{eq:def_comm}
    C_l&\equiv\{{H}^l,{F}\}\nonumber\\
    &\equiv[\underbrace{{H},[{H},...[{H},[{H}}_{\text{$l$ times}},{F}]]...]]
\end{align}
and where $i$ and $j$ are any pair of integers fulfilling $i+j=k$. By definition $C_0\equiv F$.

For odd moments, Eq.~\eqref{eq:mk_gen} can be further expressed in terms of a commutator
\begin{equation}\label{eq:mk_odd}
    m_k=\frac{1}{2}(-1)^{i}\braket{\Psi^{\sigma_0}_0|[C_i,C_j]|\Psi^{\sigma_0}_0}\,.
\end{equation}
The last step provides a useful simplification to the structure of the operator whose ground-state expectation value is to be computed. Indeed, taking $F$ to be a one-body operator, while the product $C_iC_j$ contains up to $[(n-1)k+2]$-body operators, $n$ being the highest-rank component of $H$, the commutator contains only up to $[(n-1)k+1]$-body operators. Because even moments can only  be written in terms of anti-commutators that have the same many-body rank as the product $C_iC_j$, this simplification does not occur in this case.

Eventually, two sets of \textit{moment operators} are introduced according to
\begin{subequations}\label{eq:m_operator}
    \begin{align}
        \breve{M}_k(i,j)&\equiv(-1)^{i}C_iC_j&&\forall\;\; k\geq0\,,\\
        M_k(i,j)&\equiv\frac{1}{2}(-1)^{i}[C_i,C_j]&&\forall\;\;\text{odd}\;\;k>0\,,\label{eq:odd_operator}
    \end{align}
\end{subequations}
whose expectation value in $\ket{\Psi^{\sigma_0}_0}$ delivers $m_k$. 

Based on a Hamiltonian $H$ containing up to three-body operators, the algebraic expressions of the tensors defining $M_1(1,0)$ are explicitly derived in \ref{sec:mom_expl} . The result is used to numerically compute the PGCM $m_1$ moment associated with the monopole operator $F=r^2$ via the GSEV approach in Sec.~\ref{sec:resultscomparisonmethods}.

\subsection{Alternative formulation}
\label{sec:srg_mom}

It is possible to access the operator $M_k(j+1,j)$ associated with the odd positive moment $m_k$ in an alternative way. To do so, the similarity-transformed Hamiltonian
\begin{align}\label{eq:BKH}
    H_k(\eta)&\equiv e^{-\eta G_k}He^{\eta G_k}\nonumber\\
    &=H+\eta[H,G_k]+\frac{1}{2!}\eta^2[[H,G_k],G_k]+\mathcal{O}(\eta^3)\,,
\end{align}
is introduced, where the expansion in powers of the parameter $\eta$ results from the application of Baker-Campbell-Hausdorff's identity. To match the expression given in Eq.~\eqref{eq:odd_operator} one takes $i=j+1$ and
\begin{equation}
    G_k\equiv C_j\quad\text{with  $k\equiv2j+1$}\,,
\end{equation}
such that 
\begin{equation}\label{eq:sym_op}
    M_k(j+1,j)=\frac{1}{2}(-1)^{j+1}\frac{\partial^2}{\partial\eta^2}H_k(\eta)\bigg|_{\eta=0}\,.
\end{equation}
Based on a Hamiltonian $H$ containing up to three-body operators, the algebraic expressions of the tensors defining $M_1(1,0)$ are also derived in \ref{sec:mom_expl} as a way to validate the correctness of the expressions obtained via the more direct commutator approach laid down in Sec.~\ref{MomentOP}.

\subsection{Practical merits and limitations}
\label{sec:mom_teo_app}

The great practical advantage of the GSEV approach is to access strength function's moments based on the sole knowledge of the nuclear ground state. This indeed is a tremendous simplification given that accessing a complete-enough set of excited states  constitutes a challenge within any state-of-the-art \textit{ab initio} many-body method\footnote{Interestingly, once the investment to set up a moment operator has been done, it can be employed with any method delivering the many-body ground-state.}

Such a benefit however comes at the price of evaluating the ground-state expectation value of operators (Eq.~\eqref{eq:m_operator}) whose many-body complexity increases with the moment order. The set of moment operators indeed involve the hierarchy of operators
    \begin{align*}
        C_1=&[H,F]\,,\\
        C_2=&[H,C_1]\,,\\
        &\vdots\nonumber\\
        C_{j+1}=&[H,C_j]\,,
    \end{align*}
whose many-body rank increases with $j$ due to the new commutator involved at each step. With $F$ a one-body operator and $H$ containing up to three-body operators, $C_1$ contains up to three-body operators, $C_2$ up to five-body operators, $C_3$ up to seven-body operators, i.e. $C_j$ contains up to $(2j+1))$-body operators. As a result, $\breve{M}_k(i,j)$ contains up to $(2k+2)$-body operators and $M_k(i,j)$ up to $(2k+1)$ operators. For example, $\breve{M}_0(0,0)$ contains up to two-body operators and $M_1(1,0)$ contains up to three-body operators. Knowing that dealing with three-body operators constitutes the current computational limit, it makes possible to compute both $m_0$ and $m_1$ exactly via the GSEV approach in PGCM calculations\footnote{Because in practice the Hamiltonian is first rank-reduced to be an effective two-body operator~\cite{Frosini21a}, $M_1(1,0)$ only contains up to two-body operators in present calculations.}. Moving further, $\breve{M}_2(1,1)$  contains up to six-body operators and $M_3(2,1)$ contains up to seven-body operators, which makes them beyond reach\footnote{Because the Hamiltonian is first rank-reduced to be an effective two-body operator~\cite{Frosini21a}, $\breve{M}_2(1,1)$ would only contain up to four-body operators whereas $M_3(2,1)$ would only contain up to five-body operators in the present calculations. It still makes them beyond the reach of current capabilities.}. While it is in principle possible to design approximations to $\breve{M}_2(1,1)$ and $M_3(2,1)$ based on rank-reduction techniques~\cite{Frosini21a}, this avenue is not pursued in the present work and PGCM moments such as $m_2$ and $m_3$ are accessed via the SOES approach.

\subsection{Pseudo-GSEV approach to ${m_{-1}}$}

The GSEV delivers an alternative strategy to compute moments based on the introduction of moment operators. However, this strategy does not apply to $m_k$ with $k\leq0$. The way to evaluate $m_{-1}$, which in the SOES approach reads as
\begin{equation}
m_{-1}=\sum_{\nu}\frac{|\braket{\Psi^{\sigma}_\nu|F|\Psi^{\sigma_0}_0}|^2}{E^{\sigma}_\nu-E^{\sigma_0}_0}\,, \label{SOESm-1}
\end{equation}
via a pseudo-GSEV approach relies on time-independent perturbation theory. Perturbing the system by the external field $F$, the Hamiltonian becomes
\begin{equation}
    H(\lambda)\equiv H+\lambda F \,,
\end{equation}
and the associated Schr\"odinger equation for the ground state is\footnote{Depending on the nature of the operator $F$, eigenstates of $H(\lambda)$ may or may not share the same symmetry quantum numbers $\sigma$ as those of $H$. While this is indeed the case for $F=r^2$, it is not true in general such that the superscript $\sigma_0$ is dropped in $\ket{\Psi_0(\lambda)}$.}
\begin{equation}
    H(\lambda)\ket{\Psi_0(\lambda)}=E_0(\lambda)\ket{\Psi_0(\lambda)}\,.
\end{equation}
In the small-$\lambda$ limit, perturbation theory allows one to expand $\ket{\Psi_0(\lambda)}$ in powers of $\lambda$  according to~\cite{Shavitt2009}
\begin{equation}
\ket{\Psi_0(\lambda)}=\ket{\Psi^{\sigma_0}_0}+\lambda\sum_\nu\frac{\braket{\Psi^{\sigma}_\nu|F|\Psi^{\sigma_0}_0}}{E^{\sigma_0}_0-E^{\sigma}_\nu}\ket{\Psi^{\sigma}_\nu}+\mathcal{O}(\lambda^2)\,.
\end{equation}
The variation of the ground-state expectation values of a generic operator $Q$ and of the Hamiltonian $H$ read as
\begin{subequations}
    \begin{align}
        \delta\braket{Q}\equiv&\braket{\Psi_0(\lambda)|Q|\Psi_0(\lambda)}-\braket{\Psi^{\sigma_0}_0|Q|\Psi^{\sigma_0}_0}\nonumber\\
        =&-\lambda\sum_{\nu}\biggl\{\braket{\Psi^{\sigma_0}_0|Q|\Psi^{\sigma}_\nu}\frac{1}{E^{\sigma}_\nu-E^{\sigma_0}_0}\braket{\Psi^{\sigma}_\nu|F|\Psi^{\sigma_0}_0} \nonumber\\
        &\,\,\,\,\,\,+\braket{\Psi^{\sigma_0}_0|F|\Psi^{\sigma}_\nu}\frac{1}{E^{\sigma}_\nu-E^{\sigma_0}_0}\braket{\Psi^{\sigma}_\nu|Q|\Psi^{\sigma_0}_0}\biggr\}\nonumber\\
        &+\mathcal{O}(\lambda^2)\,,\\
        \delta\braket{H}\equiv&\braket{\Psi_0(\lambda)|H|\Psi_0(\lambda)}-\braket{\Psi^{\sigma_0}_0|H|\Psi^{\sigma_0}_0}\nonumber\\
        =&\lambda^2\sum_\nu\frac{|\braket{\Psi^{\sigma}_\nu|F|\Psi^{\sigma_0}_0}|^2}{E^{\sigma}_\nu-E^{\sigma_0}_0}+\mathcal{O}(\lambda^3)\,,\label{eq:pert_H}
    \end{align}
\end{subequations}
where the term linear in $\lambda$ disappears in Eq. \eqref{eq:pert_H} due to the fact that $\{| \Psi^{\sigma}_\nu \rangle, \nu = 0, \ldots, \nu_{\text{max}} \}$ constitutes an orthonormal eigenbasis of $H$. It is easy to see that both expressions provide a direct link to $m_{-1}$ if $Q=F$, i.e.
\begin{subequations}
    \begin{align}
        m_{-1}&=-\frac{1}{2}\frac{\partial\braket{\Psi_0(\lambda)|F|\Psi_0(\lambda)}}{\partial\lambda}\bigg|_{\lambda=0}\,,\label{eq:m-1_F}\\
        m_{-1}&=\frac{1}{2}\frac{\partial^2\braket{\Psi_0(\lambda)|H|\Psi_0(\lambda)}}{\partial\lambda^2}\bigg|_{\lambda=0}\,.
    \end{align}
\end{subequations}
Notice that the first contribution to the variation of the ground-state energy is of order $\lambda^2$, which makes in general Eq. \eqref{eq:m-1_F} a more reliable numerical option to compute $m_{-1}$. While this pseudo-GSEV approach can be rather straightforwardly implemented within PGCM calculations, it is postponed to  a later work such that numerical values of $m_{-1}$ presented below actually rely on the SOES approach (Eq.~\eqref{SOESm-1}).

\section{Numerical setting}
\label{technical}

The PGCM formalism and the characteristics of the numerical applications were detailed in Paper I. In particular, the definition of the three mean-square-radius-like operators $r^2$, $r^{2}_{\text{lab}}$ and $r^{2}_{\text{int}}$ under use all throughout the present paper can be found in Paper I. All calculations presented here use the same setting as in Paper II. 

A one-body spherical harmonic oscillator basis characterised by the optimal frequency $\hbar\omega=12$~MeV is employed. All states up to $e_{\!_{\;\text{max}}}\equiv\text{max}(2n+l)=10$ are included, with $n$ the principal quantum number and $l$ the orbital angular momentum. The representation of three-body operators is further restricted by only employing three-body states  up to $e_{\!_{\;\text{3max}}}=14$.

A Hamiltonian based on chiral effective field theory ($\chi$EFT) and built at next-to-next-to-next-to-leading-order (N$^3$LO)~\cite{Hu20} is employed. It contains consistent two- (2N) and three-nucleon (3N) interactions and is further evolved via similarity renormalization group (SRG) transformations~\cite{Bogner:2009bt} to the low-momentum scale $\lambda=1.88$~fm$^{-1}$ (i.e. flow parameter $\alpha$=0.08~fm$^4$) and  truncated at the three-body operator level.  The resulting three-body force is approximated via the rank-reduction method developed in Ref.~\cite{Frosini21a}. 

Two-dimensional (2D) PGCM calculations mix a set of constrained HFB states with axial symmetry using the root-mean-square radius $r\equiv \sqrt{\langle r^2_{\text{lab}} \rangle}$ and the axial mass quadrupole deformation parameter $\beta_2$ as generator coordinates. The QRPA calculations are performed at the HFB minimum via the quasi-particle finite amplitude method (QFAM)~\cite{Beaujeault23a}. The QFAM monopole moments are computed via the contour integration of the response function in the complex energy plane~\cite{Hinohara15a}.

Eventually, the present analysis is based on the (P)GCM and QFAM monopoles responses of $^{16}$O, $^{24}$Mg, $^{28}$Si and $^{46}$Ti. Given that the present paper only focuses on spectral moments, the reader is referred to Paper II for a detailed analysis of the  corresponding strength functions.

\section{SOES and GSEV approaches to $m_1$}
\label{sec:mom_meth_comp}

\begin{table*}[t]
    \centering
    \begin{tabular}{lrlrlrlrlrl}
    	\hline
    	\hline\vspace{-0.4cm}\\
        ~ & \multicolumn{2}{c}{$^{16}$O} & \multicolumn{2}{c}{$^{24}$Mg} & \multicolumn{2}{c}{$^{28}$Si} & \multicolumn{2}{c}{$^{28}$Si$_{\;\text{iso}}$} & \multicolumn{2}{c}{$^{46}$Ti} \\ 
        ~ & SOES & GSEV & SOES & GSEV & SOES & GSEV & SOES & GSEV & SOES & GSEV \\ 
        \hline
        GCM & 7940 & 8611 & 16676 & 17850 & 21046 & 22384 & 22104 & 23625 & 43185 & 46776 \\ 
        PGCM & 8386 & 8617 & 17178 & 17978 & 21490 & 22526 & 22846 & 24016 & 44392 & 47046 \\ 
        \hline
        \hline 
    \end{tabular}
    \caption{GCM and PGCM $m_1$ monopole moments computed via the SOES and GSEV approaches. All quantities are in fm$^4$MeV.}
    \label{tab:fomula_E0_(P)GCM}
\end{table*}

\subsection{Rationale}
\label{sec:rationale}

From a formal standpoint, the equivalence between the SOES and GSEV approaches relies on the completeness assumption from Eq.~\eqref{eq:id_res_mom} allowing one to use the identity resolution. While the GSEV value of a given moment can be considered to be the formal value of reference, the SOES value is the one corresponding to the strength function actually computed in practice on the basis of a necessarily incomplete set of excited states\footnote{The moment obtained via the GSEV approach constitutes an upper bound of the value computed via the SOES approach.}. In this context, the agreement between the two values constitutes an internal-consistency test for the employed many-body method {\it relative to} the excitation operator $F$ of interest. The agreement tests whether the vector $\bold{F}| \Psi_{0}^{\sigma} \rangle$ belongs to the subspace $S$ spanned by the set of computed eigenstates $\{| \Psi^{\sigma}_\nu \rangle , \nu = 0, \ldots, \nu_{\text{max}} \}$ explicitly at play in the SOES approach.

\begin{figure}
	\centering
	\includegraphics[width=\columnwidth]{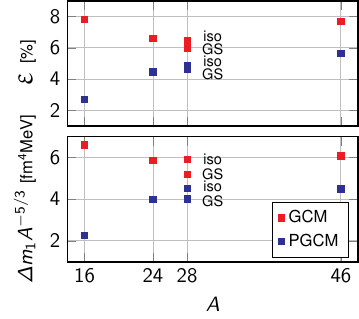}
	\caption{Difference between monopole $m_1$ values obtained via the GSEV and SOES approaches in (P)GCM calculations as a function of $A$. Upper panel: difference in percentage. Lower panel: absolute difference multiplied by $A^{-5/3}$ to remove the expected trivial $A$ dependence (see Eq.~\eqref{eq:EWSR0}).}
	\label{fig:m1_method_diff}
\end{figure}

Presently employed (P)GCM eigenstates are linear combinations of the non-orthogonal (projected) Bogoliubov vacua defining the set $\{  (P^{\sigma}) |\Phi(r^2,\beta_2) \rangle\}$ with $\,r\in[r_{\text{min}},r_{\text{max}}]$ and $\beta_2 \in[\beta_{\text{min}},\beta_{\text{max}}]$ ($\sigma \in \text{IRREPs}$). Consequently,
the (P)GCM subspace $S^{(\sigma)}$ is nothing but the span of that set. In Appendix C of Ref.~\cite{Flocard76a}, the monopole and quadrupole operators were shown to be indeed exhausted for a GCM calculation based on Slater determinants built out of the lowest eigenstates of axially deformed harmonic oscillators, the two generator coordinates being the corresponding axial and perpendicular oscillator frequencies. While realistic (P)GCM calculations rely on more general Bogoliubov vacua (and include particle-number and angular-momentum projections), such a proof gives some confidence that the monopole operator might be well exhausted in present 2D (P)GCM calculations using $r^2$ and $\beta_2$ as generator coordinates. It is the goal of the present section to test quantitatively to which extent this is indeed the case for $m_1$.

\subsection{Results}
\label{sec:resultscomparisonmethods}

\begin{table*}[]
    \centering
    \begin{tabular}{lrclrclrcl}
    	\hline
    	\hline\vspace{-0.4cm}\\
        ~ & \multicolumn{3}{c}{$^{16}$O} & \multicolumn{3}{c}{$^{24}$Mg} &  \multicolumn{3}{c}{$^{46}$Ti} \\  
        ~ & \multicolumn{3}{c}{($\beta_2=0.00$)} & \multicolumn{3}{c}{($\beta_2=0.56$)} &  \multicolumn{3}{c}{($\beta_2=0.27$)} \\  
        SOES & GCM &$\varepsilon$ [\%]& PGCM & GCM & $\varepsilon$ [\%] & PGCM & GCM & $\varepsilon$ [\%] & PGCM \\ 
        \hline
        $m_{-1}$ [fm$^4$MeV$^{-1}$] & 17.83 &0.4& 17.90 & 57.95 &0.4& 58.16 & 120.75 &0.1& 120.86 \\ 
        $m_0$    [fm$^4$]           & 369 &2.1& 377 & 955 &1.4& 969 & 2226 &1.7& 2264 \\ 
        $m_1$    [fm$^4$MeV]        & 7940 &5.2& 8386 & 16676 &2.9& 17178 & 43185 &2.7& 44392 \\ 
        $m_2$    [fm$^4$MeV$^2$]    & 182667 &11.0& 205277 & 315053 &4.4& 329629 & 871376 &3.2& 900105 \\ 
        $m_3$    [fm$^4$MeV$^3$]    & 4718706 &19.8& 5887075 & 6661187 &5.3& 7031885 & 18469397 &2.7& 18976050 \\ 
        \hline
        \hline
    \end{tabular}
    \caption{Monopole moments computed using the SOES approach for GCM and PGCM calculations of $^{16}$O, $^{24}$Mg and $^{46}$Ti. Numbers in between GCM and PGCM results indicate the variation between the former and the latter in percentage.}
    \label{tab:proj_E0_sum1}
\end{table*}

\begin{table*}[]
\centering
\begin{tabular}{lrclrclrcl}
\hline\hline
 & \multicolumn{3}{c}{$^{16}$O}                & \multicolumn{3}{c}{$^{24}$Mg}               & \multicolumn{3}{c}{$^{46}$Ti}  \\
 & \multicolumn{3}{c}{($\beta_2=0.00$)} & \multicolumn{3}{c}{($\beta_2=0.56$)} &  \multicolumn{3}{c}{($\beta_2=0.27$)} \\
SOES & GCM & $\varepsilon$ [\%] & PGCM & GCM & $\varepsilon$ [\%] & PGCM & GCM & $\varepsilon$ [\%] & PGCM  \\
\hline
$\tilde{E}_1$ & 21.10& 2.5& 21.64& 16.96& 1.3& 17.19& 18.91& 1.3& 19.17\\
$\bar{E}_1$   & 21.51& 3.4& 22.26& 17.46& 1.5& 17.72& 19.40& 1.1& 19.60\\
$\sigma$      & 5.67 &19.3&  7.02&  5.01& 1.7&  5.09&  3.89& 7.1&  3.63\\
\hline\hline
\end{tabular}
\caption{Average energies and dispersion computed using the SOES for GCM and PGCM calculations of $^{16}$O, $^{24}$Mg and $^{46}$Ti. All results are expressed in MeV units. Numbers in between GCM and PGCM results indicate the variation between the former and the latter in percentage.}
\label{tab:proj_new}
\end{table*}

The (P)GCM $m_1$ values obtained from both evaluation methods are reported in Tab.~\ref{tab:fomula_E0_(P)GCM}. Furthermore, their difference (rescaled  according to their expected $A^{5/3}$ scaling; see Eq.~\eqref{eq:EWSR0}) is displayed in Fig.~\ref{fig:m1_method_diff} along with the difference in percentage. 

Results obtained via the SOES approach are about $6-7$\% smaller than their GSEV counterpart across the five cases under consideration. The underestimation of the SOES approach is stable from $A=16$ to $A=46$ once the $A^{5/3}$ scaling has been removed. The small but systematic improvement of the PGCM over the GCM is attributed to the benefit of the symmetry restoration, i.e. symmetry contaminants are removed by the angular momentum projection (AMP) on $J=0$ such that the operator $r^2$ is better exhausted by the corresponding subspace $S_P$. 

Eventually, the operator $r^2$ is exhausted, within a few percents, by the (P)GCM subspace $S_{(P)}$. This translates into the fact that the SOES approach to $m_1$ can be safely used within a few percent uncertainty\footnote{The resulting uncertainty for a moment $m_k$ can be conjectured to increase with $k$. Indeed, the energy weight $E^k$ entering $m_k$ accentuates the importance of higher-energy states as $k$ increases while the truncation of the completeness relation in the SOES approach probably affects more this higher-energy domain. Given that $m_1$ is the highest moment that can be computed exactly within the GSEV approach, this conjecture cannot be presently tested.}.

\section{Angular-momentum projection}
\label{sec:mom_amp}


The effect of AMP on the monopole moments $m_k$, $k=-1,0,1,2,3$, evaluated via the SOES approach is presently quantified by comparing results from GCM and PGCM calculations. As seen in Tab.~\ref{tab:proj_E0_sum1}, the AMP systematically enlarges $m_k$ in a way that increases with $k$. In fact, while the increase with the moment order is rather marked in $^{16}$O, it is  limited in $^{24}$Mg and has entirely disappeared in $^{46}$Ti. Thus, and even though the range of nuclei presently tested is too limited to draw general conclusions, the impact of the AMP seems to decrease with $A$.

While the behavior of specific moments is interesting, it is more pertinent to investigate how this translates into the modification of physically-relevant quantities, e.g. the mean value and the dispersion of the monopole strength function. As visible from Tab.~\ref{tab:proj_new}, the impact of the AMP on the centroid energy $\bar{E}_1$ decreases from $3.4\%$ in  $^{16}$O to $1.5\%$ in $^{24}$Mg, and eventually down to $1.1\%$  in $^{46}$Ti. Except in $^{16}$O, where it amounts to $750$\,keV, the GMR energy shift due to AMP is thus essentially negligible. The situation is similar for $\tilde{E}_1$ that is used as an alternative to evaluate the GMR energy.

The impact on the dispersion is typically more significant. Again, the set of nuclei is too limited to draw general conclusions. Still, the dispersion varies by as much as $19.3\%$ in $^{16}$O and $7.1\%$ in $^{46}$Ti. In $^{24}$Mg, the impact of AMP on $\sigma$ is small but the strongly fragmented monopole strength is in fact significantly modified as can be seen in Paper II, which reflects the fact that $\bar{E}_1$ and $\sigma$ are anyway insufficient to characterize the behavior of the strength in such a case.

\section{Comparison to QRPA}
\label{sec:mom_correlation}

\begin{table*}[]
    \centering
    \begin{tabular}{lrlrlrlrlrl}
    	\hline
    	\hline\vspace{-0.4cm}\\
        ~ & \multicolumn{2}{c}{$^{16}$O} & \multicolumn{2}{c}{$^{24}$Mg} &  \multicolumn{2}{c}{$^{46}$Ti} & \multicolumn{2}{c}{$^{28}$Si} & \multicolumn{2}{c}{$^{28}$Si$_{\;\text{iso}}$} \\ 
        GSEV & QFAM & GCM  & QFAM & GCM & QFAM & GCM & QFAM & GCM & QFAM & GCM \\ 
        \hline
        $m_1$ [fm$^4$MeV] & 8356 & 8611 & 17478 & 17850 & 46387 & 46776 & 22080 & 22384 & 23075 & 23625 \\ 
        \hline
        \hline
    \end{tabular}
    \caption{Monopole $m_1$ moment computed via the GSEV approach for GCM and QFAM calculations of $^{16}$O, $^{24}$Mg, $^{28}$Si (ground-state and prolate isomer)  and $^{46}$Ti.}
    \label{tab:QFAM_E0_GS}
\end{table*}

\label{tab:QFAM_E0_sum1}


In Paper II, the QFAM et GCM monopole strengths of $^{16}$O, $^{24}$Mg, $^{28}$Si and $^{46}$Ti were discussed at length. The corresponding $m_1$ values calculated within the GSEV approach are compared in Tab.~\ref{tab:QFAM_E0_GS} and shown to agree to better than $3\%$ across the four nuclei under consideration, GCM values being systematically larger than QFAM ones. 

The QRPA is known to fully exhaust any one-body excitation operator in odd-$k$ moments, i.e. it can be shown analytically that, within the quasi-boson approximation, odd-$k$ QRPA moments computed with the GSEV and the SOES approaches are strictly identical, the state at play in the GSEV being the HFB ground state~\cite{Thouless61a,capelli2009dielectric,Porro_thesis}. This result demonstrates the internal consistency of the QRPA as far as strength functions are concerned. While the GCM does not strictly share this property as discussed above, the GCM ground-state is {\it necessarily} a better approximation of the exact ground state than the HFB state, such that GCM moments based on the GSEV approach are necessarily better than QRPA ones\footnote{This recalls that the capacity of a method to fully exhaust the strength of a given excitation operator is {\it not} a sufficient condition to deliver a better approximation of the exact moments than a method that does not fully exhaust it.}. This is testified by the larger values of the GCM $m_1$ moment reflecting the beneficial impact of (static) correlations associated with fluctuations of $r^2$ and $\beta_2$ leading to slightly larger GCM mean-square radii compared to HFB ones\footnote{The argument qualitatively relates to the EWSR expressing $m_1$ in terms of the ground-state mean-square matter radius.}.

\begin{figure}
    \centering
    \includegraphics[width=\columnwidth]{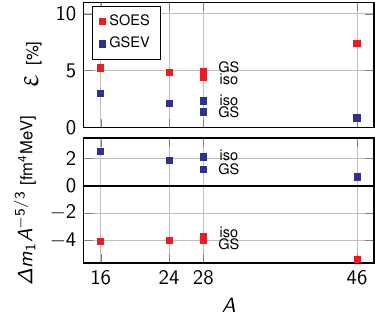}
    \caption{Difference between the GCM and the QRPA monopole $m_1$ values as a function of $A$. The GCM moment is evaluated both through the SOES and GSEV approaches. Upper panel: absolute difference in percentage. Lower panel: difference multiplied by $A^{-5/3}$ to remove the expected trivial $A$ dependence (see Eq.~\eqref{eq:EWSR0}).}
    \label{fig:m1_sum_diff}
\end{figure}

\begin{figure}[t]
    \centering
    \includegraphics[width=.8\columnwidth]{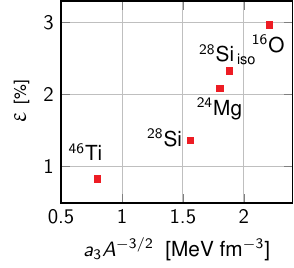}
    \caption{Relative difference between the QFAM and the GSEV GCM monopole $m_1$ values as a function of the cubic coefficient $a_3$ (see Sec.~6 of Paper II). The factor $A^{-3/2}$ is included to remove the trivial $A$ dependence of $a_3$. See text for details.}
    \label{fig:m1_anharm}
\end{figure}

The trend with $A$ of the difference between GCM and QRPA $m_1$ values is better inferred from Fig.~\ref{fig:m1_sum_diff}. Given the hypothesis at the heart of the QRPA, such a difference is expected to increase with the degree of \textit{anharmonicity} of the system. As expected, and as discussed in Sec.~6 of Paper II, larger systems are more harmonic than lighter ones. This is indeed consistent with the fact that the difference with GCM values decreases with $A$. This interpretation is further supported by Fig.~\ref{fig:m1_anharm} where the difference is shown to grow with the size of the cubic coefficient\footnote{The cubic coefficient is rescaled by $A^{-3/2}$ to remove its trivial $A$ dependence due to the use of the rms radius as the variable in the fitted function; see Paper II for details.} $a_3$ extracted in Sec.~6 of Paper II.

Figure~\ref{fig:m1_sum_diff} also displays the deviation of QFAM $m_1$ values from GCM results based on the SOES approach. In this case, QFAM values are systematically a few percent {\it above} GCM ones; i.e. they are located in between the two sets of GCM values. Eventually, the disagreement between QRPA and GCM is smaller than the uncertainty in the evaluation of  the GCM values. Contrary to values based on the GSEV approach, GCM values obtained from the SOES approach do not converge towards QRPA  as $A$ increases and thus do not scale as expected with the harmonic character of the system.

\begin{table*}[]
\centering
\begin{tabular}{lrclrclrcl}
\hline
\hline
 & \multicolumn{3}{c}{$^{16}$O}       & \multicolumn{3}{c}{$^{24}$Mg}      & \multicolumn{3}{c}{$^{46}$Ti}      \\
 & \multicolumn{3}{c}{($\beta_2=0.00$)} & \multicolumn{3}{c}{($\beta_2=0.56$)} &  \multicolumn{3}{c}{($\beta_2=0.27$)} \\
 & QFAM  & $\varepsilon$ [\%] & GCM   & QFAM  & $\varepsilon$ [\%] & GCM   & QFAM  & $\varepsilon$ [\%] & GCM   \\
\hline
$\bar{E}_1$   & 22.33 & 3.8  & 21.51 & 18.48 & 5.9  & 17.46 & 20.15 & 3.9 & 19.40 \\
$\sigma$      & 5.55  & 2.1  & 5.67  & 5.58  & 11.5 & 5.01  & 3.96  & 1.8 & 3.89 \\
\hline
\hline
\end{tabular}
\caption{Centroid energy and dispersion from QFAM and GCM(SOES) calculations of $^{16}$O, $^{24}$Mg and $^{46}$Ti. All results are expressed in MeV units. Numbers in between QFAM and GCM results indicate the variation between the former and the latter in percentage.}
\label{tab:corr_new}
\end{table*}

Eventually, the centroid and the dispersion of the QRPA monopole strength function are compared to GCM values based on the SOES approach in Tab.~\ref{tab:corr_new}. The GCM centroid energy is typically $4-6\%$ below the QRPA one\footnote{While it is true for the centroid of the actually computed GCM strength function (SOES value), the formal (GSEV) value not computed here is probably {\it higher} in view of the behavior of $m_1$ studied above.}, which amounts to less than $1$\,MeV difference in the studied nuclei. The QRPA and GCM dispersions are also very consistent, especially in view of the remaining many-body uncertainty.

\section{Energy weighted sum rule}
\label{sec:ewsr_ex}

\subsection{Definition}
\label{sec:ewsr}

The EWSR is a standard quantity in GRs studies and provides a good indicator of the degree of collectivity of nuclear excitations. Furthermore, the EWSR is used to extract strength functions from experimental data as briefly recalled in \ref{sec:relation_obs}. 

The ESWR relies on an analytical evaluation of $m_1$ via Eq.~\eqref{eq:m_operator}, i.e. using the GSEV approach. Targeting the first moment of the isoscalar monopole strength function, the similarity transformation of the Hamiltonian in Eq.~\eqref{eq:BKH} computed with the isoscalar local operator $G_0=F=r^2$ is nothing but a local-gauge transformation.  In case inter-nucleon interactions are local-gauge invariant, they do not contribute to the quadratic term in $\eta$ in Eq.~\eqref{eq:BKH} which is nothing but the operator $M_1(1,0)$ delivering $m_1$. Under such an assumption only the (laboratory-frame) kinetic-energy operator 
\begin{equation}
\label{eq:t_lab}
T_{\text{lab}} \equiv \frac{1}{2m} \sum_{i=1}^{A} p^2_i \, ,
\end{equation}
contributes to $M_1(1,0)$ such that $m_1$ is obtained analytically under the form~\cite{Bohigas79a,Hinohara:2019ych}
\begin{equation}
    \text{EWSR}_{\text{lab}}(r^2)=\frac{2\hbar^2A}{m} \langle \Psi^{\sigma_0}_{0} |r^2_{\text{lab}} | \Psi^{\sigma_0}_{0} \rangle \, , \label{eq:EWSR0}
\end{equation}
which constitutes the textbook EWSR formula for the isocalar monopole mode. Interestingly,  $\text{EWSR}_{\text{lab}}(r^2)$ is proportional to the (laboratory-frame) ground-state mean-square matter radius. Thus, accessing it only requires the computation of that mean-square radius from the many-body method of interest, e.g. using the Bogoliubov state at the HFB minimum in QFAM calculations or the (P)GCM ground state in (P)GCM calculations. 

However, it happens that nuclear excitations of interest are \textit{intrinsic} excitations. Consequently, present many-body calculations employ the intrinsic Hamiltonian $H$ containing the intrinsic kinetic-energy operator $T_{\text{int}}\equiv T_{\text{lab}}-T_{\text{cm}}$, with the subtracted center-of-mass kinetic-energy operator reading as
\begin{equation}
\label{eq:t_cm}
T_{\text{cm}} \equiv \frac{1}{2mA} \sum_{i,j=1}^{A} \vec{p}_i \cdot \vec{p}_j \, .
\end{equation}
In this context, the monopole $m_1$ moment reads as
\begin{align}
	\label{eq:split_mom}
	m_1&=\frac{1}{2} \langle \Psi^{\sigma_0}_{0} | [r^2,[H,r^2]] | \Psi^{\sigma_0}_{0} \rangle \nonumber\\
	&=\text{EWSR}_{\text{lab}}(r^2)-\frac{1}{2} \langle \Psi^{\sigma_0}_{0} | [r^2,[T_{\text{cm}},r^2]]| \Psi^{\sigma_0}_{0} \rangle \nonumber\\
	&\equiv\text{EWSR}_{\text{lab}}(r^2) -\frac{2\hbar^2A}{m}  \langle \Psi^{\sigma_0}_{0} |R^2_{\text{cm}} | \Psi^{\sigma_0}_{0} \rangle  \,,
\end{align}
with $\vec{R}_{\text{cm}}$ the center-of-mass position vector. The derivation of the correction term from $T_{\text{cm}}$ is provided in \ref{correctCMEWSR}. Eventually, the two terms can be combined such that the appropriate, i.e. \textit{intrinsic}, ESWR is given by 
\begin{align}
    \text{EWSR}_{\text{int}}(r^2)&= \frac{2\hbar^2A}{m} \langle \Psi^{\sigma_0}_{0} | r^2_{\text{int}}| \Psi^{\sigma_0}_{0} \rangle  \, , \label{finalEWSRintbulk}
\end{align}
and thus amounts to using the intrinsic mean-square radius rather than the laboratory-frame one. 

The $\text{EWSR}_{\text{int}}(r^2)$ must in principle be fulfilled in \textit{ab initio} calculations given that $\chi$EFT-based 2N and 3N interactions are meant to be local-gauge invariant, which is a necessary condition to achieve a consistent coupling to the electromagnetic field~\cite{Krebs:2016rqz}. However, enforcing the local-gauge invariance is not straightforward in practice. First, it cannot be exactly fulfilled if the same EFT truncation level is applied to both nuclear interactions and currents, even in the case of dimensional regularization. Second, the use of (nonlocal) cutoff regulators makes its fulfillment even more challenging~\cite{Marcucci:2004cm}. Eventually, existing $\chi$EFT-based 2N and 3N interactions are not strictly local-gauge invariant and it is our goal to quantify such a feature by testing the exhaustion of $\text{EWSR}_{\text{int}}(r^2)$ by the computed $m_1$. 

The potential breaking of the local gauge invariance can be straightforwardly formulated by schematically expressing the intrinsic Hamiltonian as
\begin{align}
	H &\equiv T_{\text{lab}} - T_{\text{cm}} + V   \nonumber \\
 &= T_{\text{lab}} - T_{\text{cm}} + V_{\text{lgi}} + V_{\cancel{\text{lgi}}} \,, \label{schematicH}
\end{align}
where $V_{\cancel{\text{lgi}}} \equiv V - V_{\text{lgi}} $ formally defines the departure of the nuclear interactions from their local-gauge invariant formulation. Given Eq.~\eqref{schematicH}, the monopole $m_1$ moment effectively reads in practice as
\begin{align}
	m_1&=\frac{1}{2}\braket{[r^2,[H,r^2]]}_{\text{gs}} \nonumber\\
	&=\text{EWSR}_{\text{int}}(r^2)+\frac{1}{2}\braket{[r^2,[V_{\cancel{\text{lgi}}},r^2]]}_{\text{gs}}\nonumber\\
	&\equiv \text{EWSR}_{\text{int}}(r^2) + \delta m^{\cancel{\text{lgi}}}_{1} \,,
\end{align}
where $\delta m^{\cancel{\text{lgi}}}_{1}$ quantifies the effective breaking of $\text{EWSR}_{\text{int}}(r^2)$.

\subsection{$\text{EWSR}_{\text{lab}}$ versus $\text{EWSR}_{\text{int}}$}
\label{sec:ewsr_results}

\begin{table}
    \centering
    \begin{tabular}{l r r }
    	\hline
    	\hline\vspace{-0.4cm}\\
        $\text{EWSR}_{\text{int}}$  & QFAM  & PGCM  \\ 
        \hline
        $^{16}$O                    & 8462  & 8475  \\
        $^{24}$Mg                   & 17480 & 17658 \\
        $^{28}$Si                   & 22004 & 22129 \\
        $^{28}$Si$_{\;\text{iso}}$  & 22883 & 23352 \\ 
        $^{46}$Ti                   & 45600 & 45844 \\
        \hline
        \hline
    \end{tabular}
    \caption{Isoscalar monopole $\text{EWSR}_{\text{int}}$ from Eq.~\eqref{finalEWSRintbulk} for QFAM and PGCM calculations of $^{16}$O, $^{24}$Mg, $^{28}$Si and $^{46}$Ti. All results are expressed in fm$^4$MeV.}
    \label{tab:EWSR0}
\end{table}

\begin{table}
    \centering
    \begin{tabular}{l r r }
    	\hline
    	\hline\vspace{-0.4cm}\\
        $\text{EWSR}_{\text{lab}}$  & QFAM  & PGCM  \\ 
        \hline
        $^{16}$O                    &  8832 &  8851 \\
        $^{24}$Mg                   & 17885 & 18064 \\
        $^{28}$Si                   & 22415 & 22540 \\
        $^{28}$Si$_{\;\text{iso}}$  & 23287 & 23749 \\ 
        $^{46}$Ti                   & 46053 & 46299 \\
        \hline
        \hline
    \end{tabular}
    \caption{Isoscalar monopole $\text{EWSR}_{\text{lab}}$from Eq.~\eqref{eq:EWSR0} for QFAM and PGCM calculations of $^{16}$O, $^{24}$Mg, $^{28}$Si and $^{46}$Ti. All results are expressed in fm$^4$MeV.}
    \label{tab:EWSR0_lab}
\end{table}

\begin{table}[]
\centering
\begin{tabular}{lrrrr}
\hline
\hline\vspace{-0.3cm}\\
         & \multicolumn{2}{c}{$\sqrt{\braket{r^2_{\text{lab}}}_{\text{gs}}}$} & \multicolumn{2}{c}{$\sqrt{\braket{r^2_{\text{int}}}_{\text{gs}}}$} \\
         \hline
         & HFB         & PGCM        & HFB         & PGCM        \\
         \hline
    $^{16}$O                    & 2.580       & 2.583       & 2.525       & 2.527       \\
    $^{24}$Mg                   & 2.998       & 3.013       & 2.963       & 2.979       \\
    $^{28}$Si                   & 3.107       & 3.115       & 3.078       & 3.087       \\
    $^{28}$Si$_{\;\text{iso}}$  & 3.167       & 3.198       & 3.139       & 3.171       \\
    $^{46}$Ti                   & 3.474       & 3.484       & 3.457       & 3.467      \\
\hline
\hline
\end{tabular}
\caption{Ground-state expectation value of point-matter nuclear radii in the laboratory and intrinsic frames for HFB and PGCM calculations. All results are expressed in fm.}
\label{tab:r_lab_vs_int}
\end{table}

Values of $\text{EWSR}_{\text{int}}(r^2)$ ($\text{EWSR}_{\text{lab}}(r^2)$) from QFAM and PGCM calculations are reported in Tab.~\ref{tab:EWSR0} (Tab.~\ref{tab:EWSR0_lab}). While PGCM values are systematically larger, the difference is eventually very small. These features reflect the behavior of the point-matter radii reported in Tab.~\ref{tab:r_lab_vs_int}\footnote{The difference between HFB and PGCM radii relates to the impact of so-called static correlations beyond the mean-field included into the PGCM ansatz. In general, static correlations have little impact on radii, the exceptions being light spherical nuclei in which they can non-negligibly increase radii and transitional nuclei in which they can strongly reduce them. In nuclei displaying a sharp total energy surface around the deformed HFB minimum, as is the case here, the impact of static correlations on the mean-square radius is typically very small~\cite{Bender06a}. The presently employed $\chi$EFT Hamiltonian typically delivers good radii such that the further addition of missing dynamical correlations to the PGCM ansatz~\cite{Frosini21a} is expected to enlarge radii.}.

\begin{figure}
    \centering
    \includegraphics[width=\columnwidth]{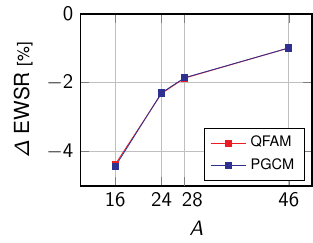}
    \caption{Relative difference between the monopole $\text{EWSR}_{\text{int}}$ and $\text{EWSR}_{\text{lab}}$ as a function of $A$ for QFAM and PGCM calculations.}
    \label{fig:EWSR_diff}
\end{figure}

The relative difference between $\text{EWSR}_{\text{int}}$ and $\text{EWSR}_{\text{lab}}$ is plotted as a function of $A$ in Fig.~\ref{fig:EWSR_diff}. Results are identical for QFAM and PGCM calculations. The $A$ dependence of the difference relates to the $-A^{-1}$ scaling driven by the center-of-mass-correction entering $\text{EWSR}_{\text{int}}$, as is analytically demonstrated in \ref{correctCMEWSR} for its one-body component. Again, the trend reflects directly how the difference of the mean-square point matter radii computed in the intrinsic frame and in the laboratory frame decreases with $A$.

\subsection{Exhaustion of Sum Rules}

\begin{figure}
    \centering
    \includegraphics[width=\columnwidth]{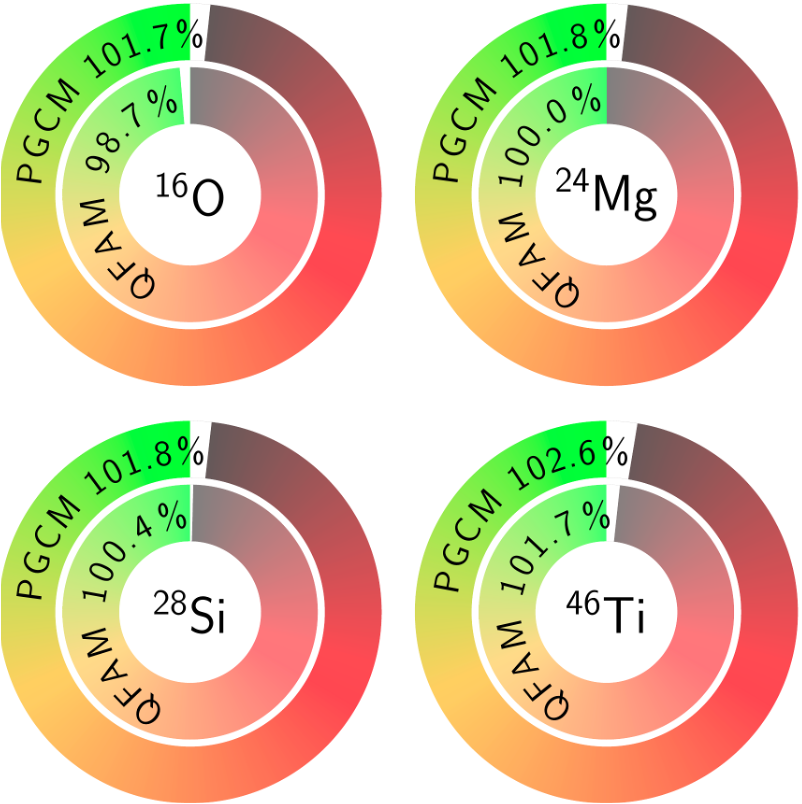}
    \caption{Exhaustion of the monopole $\text{EWSR}_{\text{int}}$  in PGCM and QFAM calculations.}
    \label{fig:fancy_mom_BE0}
\end{figure}

The actual exhaustion of the values provided in Tab.~\ref{tab:EWSR0} is now tested. Introducing the deviation in percentage
\begin{equation}
	\label{eq:quant_anharm}
	\varepsilon \equiv\left(\frac{m_1}{\text{EWSR}_{\text{int}}}-1\right)\times 100 \, ,
\end{equation}
the exhaustion $(100+\varepsilon)$ of $\text{EWSR}_{\text{int}}(r^2)$
is reported in Tab.~\ref{tab:EWSR_fomula_mono} and Fig.~\ref{fig:fancy_mom_BE0} for QFAM and PGCM calculations. It is observed that PGCM results overshoot the $\text{EWSR}_{\text{int}}(r^2)$ by $[1.7,2.6]\%$ whereas QFAM results are slightly lower and exhaust it within $[-1.3,+0.8]\%$.

Overall, the violation of $\text{EWSR}_{\text{int}}(r^2)$ due to the breaking $\delta m^{\cancel{\text{lgi}}}_{1}$ of local-gauge invariance by the presently employed $\chi$EFT interactions is small and remains below $3\%$ in the present calculations. Still, it manifests slightly differently depending on the (approximate) many-body method, the nucleus or the eigenstate under consideration. 

\begin{figure}
    \includegraphics[width=\columnwidth,right]{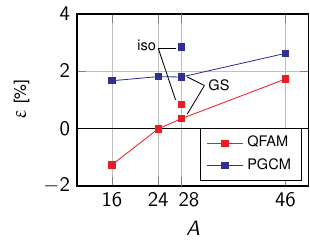}
	\caption{Percent variation of the computed monopole $m_1$ compared to the corresponding $\text{EWSR}_{\text{int}}$ as a function of $A$ for QFAM and PGCM calculations. The PGCM values are based on the GSEV approach.}
	\label{fig:nonloc}
\end{figure}

To illustrate this point more transparently, $\varepsilon$ is plotted in Fig.~\ref{fig:nonloc} as a function of $A$. The difference to $\text{EWSR}_{\text{int}}(r^2)$ evolves with $A$ for the ground states under consideration and is systematically larger for PGCM than for QRPA. One observes that the trend with $A$ is flatter for the PGCM and that QFAM results seem to approach PGCM ones as the mass increases. One may conjecture that this is a sign of better convergence of the PGCM ground-state.

Eventually, a thorough investigation of the violation of $\text{EWSR}_{\text{int}}(r^2)$ requires a larger set of nuclei and excited states as well as to employ an expansion many-body method at various truncation orders. Furthermore, $\delta m^{\cancel{\text{lgi}}}_{1}$ must be studied as a function of the chiral order and for various regularizations of the employed $\chi$EFT interactions. Such a systematic study is left to a future work.

\begin{table}
    \centering
    \begin{tabular}{l r r }
    	\hline
    	\hline\vspace{-0.4cm}\\
        \text{\% of} $\text{EWSR}_{\text{int}}$ & QFAM  & PGCM  \\ 
        \hline
        $^{16}$O                    &  98.74 & 101.67 \\
        $^{24}$Mg                   &  99.99 & 101.81 \\
        $^{28}$Si                   & 100.35 & 101.80 \\
        $^{28}$Si$_{\;\text{iso}}$  & 100.84 & 102.85 \\ 
        $^{46}$Ti                   & 101.73 & 102.62 \\
        \hline
        \hline
    \end{tabular}
    \caption{Exhaustion of the monopole $\text{EWSR}_{\text{int}}$ in PGCM and QFAM calculations.}
    \label{tab:EWSR_fomula_mono}
\end{table}

\section{Nuclear incompressibility}
\label{sec:K}

The monopole breathing mode probes the compressibility of nuclear matter. Consequently, the infinite matter incompressibility modulus $K_{\infty}$ has been extracted based on microscopic calculations of $E_{\small\text{\,GMR}}$, typically within the frame of the nuclear energy density functional (EDF) method~\cite{Garg2018a}. As a matter of fact, the extraction procedure is not unambiguous in itself. Furthermore, while originally applying it to a couple of doubly closed-shell nuclei ($^{208}$Pb and $^{92}$Zr) led to consistent values of $K_{\infty}$, the more recent use of open-shell nuclei produced conflicting results. 

The goal is to extract the value of $K_{\infty}$ associated with $\chi$EFT-based interactions via \text{ab initio} calculations. In EDF calculations, it has become customary to extract $K_{\infty}$  by computing directly the symmetric nuclear matter EOS, while checking that $E_{\small\text{\,GMR}}$ is well reproduced in a selected set of finite nuclei on the basis of the same EDF parameterization. Another approach, presently in use, consists of extracting $K_{\infty}$ from the leptodermous expansion of the finite-nucleus compressibility modulus computed microscopically~\cite{Blaizot80a}. While the former approach typically carries smaller uncertainties, the latter bypasses the need to compute the infinite matter EOS. 

The second approach was recently employed to extract  $K_{\infty}$ for NNLO$_{\text{sat}}$~\cite{Ekstrom15a} and NNLO$_{\text{opt}}$~\cite{Ekstrom:2013kea} $\chi$EFT-based Hamiltonians via symmetry-adapted no core shell model (SA-NCSM) calculations of $^{4}$He, $^{16}$O, $^{20}$Ne and $^{40}$Ca~\cite{Burrows23a}. The extracted result for NNLO$_{\text{sat}}$ ($K_{\infty}=297$) was shown to be consistent, within the rather large extrapolation uncertainties, with the value  ($K_{\infty}=253$) based on the computation of the EOS with the same Hamiltonian.

Following the same protocol but only relying on a set of intrinsically-deformed nuclei, i.e. $^{24}$Mg, $^{28}$Si and $^{46}$Ti, the compressibility modulus $K_{\infty}$ associated with the N$^3$LO Hamiltonian under use~\cite{Hu20} is presently estimated based on PGCM and QRPA calculations.  

\subsection{Finite-nucleus compression modulus}
\label{sec:KA}

The first step consists of accessing the finite-system compression modulus given by~\cite{Blaizot80a}
\begin{equation}
\label{eq:K_A}
    K_A \equiv \frac{m}{\hbar^2} \, \langle \Psi^{\sigma_0}_{0} |r^2_{\text{lab}} | \Psi^{\sigma_0}_{0} \rangle \, E^2_{\small\text{\,GMR}}\,,
\end{equation}
which thus requires the ground-state mean-square matter radius and the GMR energy as inputs. In finite, especially light and deformed, nuclei the GMR strength is not concentrated into a single peak. Consequently, the choice of $E_{\small\text{\,GMR}}$ to be used in Eq.~\eqref{eq:K_A} is neither unique nor obvious. Specific derivations support the use of $\tilde{E_1}$ or $\tilde{E}_3$ whereas general arguments also motivate the use of the centroid energy $\bar{E_1}$~\cite{Blaizot80a}.  In the following, all three cases are tested\footnote{Whenever a single mode exhausts the complete monopole response, the three energies are the same and the choice is thus straightforward.}. 

\begin{figure}
    \includegraphics[width=1.05\columnwidth,right]{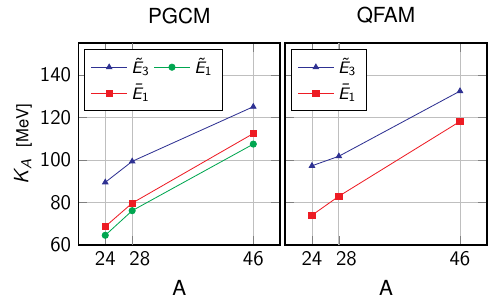}
    \caption{Finite-nuclei compression modulus $K_A$ as a function of $A$ for PGCM and QFAM calculations. Different definitions of the average GMR energy $E_{\small\text{ GMR}}$ entering Eq.~\eqref{eq:K_A} are used, see Eqs.~\eqref{eq:ave_ene_new} for the notation.}
    \label{fig:KA_vs_A}
\end{figure}

\begin{figure*}
    \centering
    \includegraphics[width=\textwidth]{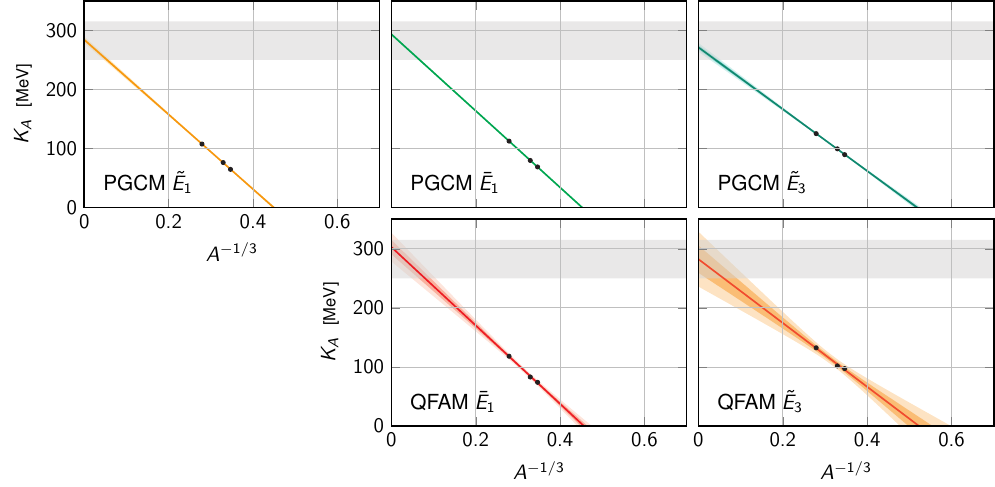}
    \caption{Finite-nucleus compression modulus $K_A$ as a function of $A^{-1/3}$ obtained from PGCM and QRPA calculations (black circles). The best fit is shown in all cases with the corresponding 1$\sigma$ (darker shade) and 2$\sigma$ (lighter shade) bands of regression (see, e.g., Chap. 3 of Ref.~\cite{draper1998applied}). The shaded gray area represents the empirically accepted range $250<K_\infty<315$~MeV~\cite{Stone14a}.}
    \label{fig:K_A}
\end{figure*}

Based on the GMR energies provided in Tab.~\ref{tab:E_GMR}, the set of $K_A$ values are given in Tab.~\ref{tab:K_A} and displayed in Fig.~\ref{fig:KA_vs_A} as a function of $A$. The higher values of $K_A$ in QRPA than in PGCM reflects the characteristics of the GMR energies pointed out earlier on whenever computing PGCM moments via the SOES approach as presently done. The spread of $K_A$ values depending on the definition of $E_{\small\text{\,GMR}}$ is the manifestation that $\tilde{E}_1$ ($\tilde{E}_3$) is more sensitive to the part of the strength located at  lower (higher) energies than $\bar{E}_1$. Eventually, $K_A$ can typically vary by as much as $30\%$ in $^{24}$Mg depending on that choice. However, this variation quickly decreases with $A$ to reach $14\%$ in $^{46}$Ti. Such a trend is encouraging in view of extracting $K_{\infty}$.

\begin{table}[]
\centering
\begin{tabular}{l|cc|ccc}
\hline
\hline\vspace{-0.4cm}\\
\multirow{2}{*}{$E_{\text{GMR}}$} & \multicolumn{2}{c|}{QFAM} & \multicolumn{3}{c}{PGCM} \\
     & \multicolumn{1}{c}{$\bar{E}_1$} & $\tilde{E}_3$ & \multicolumn{1}{c}{$\tilde{E}_1$} & \multicolumn{1}{c}{$\bar{E}_1$} & \multicolumn{1}{c}{$\tilde{E}_3$} \\
     \hline
$^{24}$Mg                & \multicolumn{1}{l}{18.48}       & 21.19         & \multicolumn{1}{l}{17.19}         & \multicolumn{1}{l}{17.72}       & \multicolumn{1}{l}{20.23}      \\
$^{28}$Si                & \multicolumn{1}{l}{18.88}       & 20.91         & \multicolumn{1}{l}{18.04}         & \multicolumn{1}{l}{18.45}       & \multicolumn{1}{l}{20.60}             \\
$^{46}$Ti                & \multicolumn{1}{l}{20.15}       & 21.33         & \multicolumn{1}{l}{19.17}         & \multicolumn{1}{l}{19.60}       & \multicolumn{1}{l}{20.68}            \\
\hline
\hline
\end{tabular}
\caption{Average GMR energies in MeV computed from QFAM and PGCM calculations according to Eqs.~\eqref{eq:ave_ene}.}
\label{tab:E_GMR}
\end{table}

\begin{table}[]
\centering
\begin{tabular}{l|rr|rrr}
\hline
\hline\vspace{-0.4cm}\\
\multirow{2}{*}{$K_A$} & \multicolumn{2}{c|}{QFAM} & \multicolumn{3}{c}{PGCM} \\
     & \multicolumn{1}{c}{$\bar{E}_1$} & $\tilde{E}_3$ & \multicolumn{1}{c}{$\tilde{E}_1$} & \multicolumn{1}{c}{$\bar{E}_1$} & \multicolumn{1}{c}{$\tilde{E}_3$} \\
\hline
$^{24}$Mg  & \multicolumn{1}{r}{74.0} & 97.3 & \multicolumn{1}{r}{64.6} & \multicolumn{1}{r}{68.7} & \multicolumn{1}{r}{89.5}\\
$^{28}$Si  & \multicolumn{1}{r}{83.0} & 101.8 & \multicolumn{1}{r}{76.2} & \multicolumn{1}{r}{79.7} & \multicolumn{1}{r}{99.4}\\
$^{46}$Ti  & \multicolumn{1}{r}{118.2} & 132.4 & \multicolumn{1}{r}{107.5} & \multicolumn{1}{r}{112.5} & \multicolumn{1}{r}{125.1} \\
\hline
\hline
\end{tabular}
\caption{Finite-nucleus compression modulus $K_A$ computed from QFAM and PGCM calculations. Values are categorised according to the definition of the GMR energy (see Eqs.~\eqref{eq:ave_ene}) employed to compute $K_A$ via Eq.~\eqref{eq:K_A}.}
\label{tab:K_A}
\end{table}

\subsection{Extraction of $K_{\infty}$}
\label{sec:Kinfty}

The method to extract $K_{\infty}$ is based on the leptodermous expansion of $K_A$ given by~\cite{Blaizot80a}
\begin{equation}
\label{eq:fit}
K_A=K_{\text{vol}}+K_{\text{surf}}A^{-1/3}+K_{\text{Coul}}Z^2A^{-4/3}+K_{\text{sym}}\beta^2\,,
\end{equation}
where $K_{\text{vol}}$, $K_{\text{surf}}$, $K_{\text{Coul}}$ and $K_{\text{sym}}$ are the volume, surface, Coulomb and symmetry contributions to the compression modulus, respectively. The parameter $\beta$ characterizes the isospin asymmetry
\begin{equation}
    \beta\equiv\frac{N-Z}{N+Z}\,,
\end{equation}
where $N$ ($Z$) denotes the neutron (proton) number. Equation~\eqref{eq:fit} is fitted based on the values of $K_A$ given in Tab.~\ref{tab:K_A} and $K_{\text{vol}}$ is interpreted as the infinite nuclear matter incompressibility $K_\infty$. Given that the Coulomb and symmetry terms do not significantly impact the asymptotic behaviour of $K_A$ for very large $A$~\cite{Burrows23a}, $K_\infty$ can be obtained via a simple linear fit in the variable $x\equiv A^{-1/3}$
\begin{equation}
\label{eq:lin_fit}
    K_A=K_\infty + K_{\text{surf}} \, x\,.
\end{equation}
While the linear fits are displayed in Fig.~\ref{fig:K_A}, the corresponding values of $K_\infty$ and $K_{\text{surf}}$ are reported in Tab.~\ref{tab:fit_param} along with the uncertainties associated with the fit. The extracted incompressibility is $K_{\infty} \approx 290$. While QRPA central values are a few MeV higher than  PGCM ones, they only differ by about $3.3\%$ and $4.2\%$ when using $E_{\small\text{\,GMR}}\equiv \bar{E}_1$ and $E_{\small\text{\,GMR}}\equiv \tilde{E}_3$, respectively. Eventually, QRPA and PGCM values are consistent within extrapolation uncertainties, which are significantly larger for QRPA than for PGCM results\footnote{The tiny extrapolation uncertainty of the PGCM results might be accidental, i.e. it may simply reflect the small number of points employed in the fit rather than a genuine behavior following strictly the $A^{-1/3}$ law of Eq.~\eqref{eq:lin_fit}. The exercise needs to be repeated in the future with a significantly larger number of points.}. 

Interestingly, while the hierarchy $K_A(\tilde{E}_1) < K_A(\bar{E}_1) < K_A(\tilde{E}_3) $ is systematically valid  for all computed nuclei with $A\leq 46$, the trends are such that the extrapolation to very large $A$ values leads to $K_{\infty}$ being the smallest for $E_{\small\text{\,GMR}}\equiv \tilde{E}_3$. Eventually, the nuclear matter incompressibility varies by $6.6\%$ ($7.5\%$) for QRPA (PGCM) between the two extreme values obtained for $\tilde{E}_3$ and $\bar{E}_1$. This confirms the trend observed above for $K_A$ as a function of $A$. 

In Fig.~\ref{fig:K_A}, the shaded gray area accounts for the generally accepted range $250<K_\infty<315$~MeV~\cite{Stone14a}. All values of $K_\infty$ fall, within extrapolation uncertainties, into this region\footnote{Values of $K_{\text{surf}}$ are also in qualitative agreement with systematic studies~\cite{Stone14a}.}. 

Uncertainties of the present theoretical predictions (partially) evaluated in Paper I are not presently propagated to $K_\infty$. While they are not negligible, they are typically subleading compared to the uncertainties associated with the choice of $E_{\small\text{\,GMR}}$ and with the extrapolation based on the leptodermous expansion. While the range of masses presently used in the fit allows one to make quantitative statement, the use of (much) heavier systems in the future will help reducing  the extrapolation uncertainty and ensure the stability of the fit. In any case, and as already stipulated in Ref.~\cite{Burrows23a}, the present work demonstrates that extrapolating the finite-nuclei compressibility modulus for a large enough set of nuclei can be complementary to the computation of the EOS in order to extract the nuclear matter compressibility. 

\begin{table}[]
    \centering
    \begin{tabular}{lll}
    \hline\hline
         & $K_\infty$ & $K_\text{surf}$ \\ \hline
        QFAM $\bar{E}_1$ & 303(12) & -664(38) \\ 
        QFAM $\tilde{E}_3$ & 283(23) & -540(72) \\ \hline
        PGCM $\tilde{E}_1$ & 284(3) & -632(9) \\ 
        PGCM $\bar{E}_1$ & 293(1) & -649(3) \\ 
        PGCM $\tilde{E}_3$ & 271(4) & -523(12) \\ \hline\hline
    \end{tabular}
    \caption{Fitting parameters for the linear extrapolation given in Eq.~\eqref{eq:lin_fit}. All quantities are expressed in MeV.}
    \label{tab:fit_param}
\end{table}

\section{Conclusions}
\label{sec:concl}

The present paper focused on the ab initio computation of the monopole strength's moments $m_k$. As a first step, the formal capacity to compute low-order moments in PGCM calculations via the ground-state expectation value of moment operators was achieved. This development was then exploited to validate the use, within a few percent uncertainty, of the approach based on the explicit sum over excited states for the first moment $m_1$ in $^{16}$O, $^{24}$Mg, $^{28}$Si and $^{46}$Ti.

With this at hand, the angular momentum projection was shown to have little impact on the centroid but to affect significantly the dispersion of the monopole strength distribution. Next, the centroid energy obtained in GCM calculations was demonstrated to be typically $4-6\%$ below  QRPA results, which amounts to less than $1$\,MeV difference in the nuclei under study. The QRPA and GCM dispersions were seen to be also very consistent.

The next part of the study focused on the EWSR and first demonstrated that its textbook expression must be corrected for the fact that nuclear excitations of interest are \textit{intrinsic} excitations in the center-of-mass frame. Having derived the appropriate {\it intrinsic} analytical EWSR, its exhaustion was shown to be violated on the level of $3\%$ as a result of the (unwanted) local-gauge symmetry breaking of the  employed $\chi$EFT-based Hamiltonian~\cite{Hu20}. 

Eventually, the finite-nucleus compressibility $K_A$ was computed in $^{24}$Mg, $^{28}$Si and $^{46}$Ti in order to extract the infinite matter nuclear incompressibility $K_{\infty}=290(15)$ that happens to be consistent, within uncertainties, with empirical expectations.

\section*{Acknowledgements}
The authors thank G. Col\`o, E. Epelbaum and U. van Kolck for useful discussions. Calculations were performed by using HPC resources from GENCI-TGCC (Contract No. A0130513012). A.P. was supported by the CEA NUMERICS program, which has received funding from the European Union's Horizon 2020 research and innovation program under the Marie Sk{\l}odowska-Curie grant agreement No 800945. A.P. and R.R. are supported by the Deutsche Forschungsgemeinschaft (DFG, German Research Foundation) – Projektnummer 279384907 – SFB 1245. R.R. acknowledges support though the BMBF Verbundprojekt 05P2021 (ErUM-FSP T07, Contract No. 05P21RDFNB).

\section*{Data Availability Statement}
This manuscript has no associated data or the data will not be deposited.

\appendix

\section{Schwartz' inequalities}
\label{sec:schwartz}

With $\rho(E)$ a positive definite function and, thus, $\rho(E)dE$ a positive measure, Schwartz's inequality reads as
\begin{equation}\label{eq:Schwartz}
    \int f^2(E)\rho(E)dE\int g^2(E)\rho(E)dE\geq\bigg(\int f(E)g(E)\rho(E)dE\bigg)^2\,,
\end{equation}
with $f$ and $g$ two arbitrary functions. The strength function $S(E)$ is defined for positive values of $E$ (see Eq. \eqref{eq:strength}), so that
\begin{equation}
    \rho(E)\equiv E^kS(E)
\end{equation}
is positive definite. For $f(E)=E$ and $g(E)=1$ Eq.~\eqref{eq:Schwartz} reads
\begin{equation}
    m_{k+2}m_k\geq m_{k+1}^2\,,
\end{equation}
which provides the sequence of inequalities in Eq. \eqref{eq:ineq}.

\section{Commutator approach}
\label{sec:momentderivation}

The introduction of moment operators first relies on expressing the moments in terms of the commutators $C_l$. This step relies on rewriting Eq. \eqref{eq:def_init}  as
\begin{align}\label{eq:dim_pass}
m_k&=\braket{\Psi^{\sigma_0}_0|\textbf{F}\textbf{H}^k\textbf{F}|\Psi^{\sigma_0}_0}\nonumber\\
&=\braket{\Psi^{\sigma_0}_0|\textbf{F}\textbf{H}^i\textbf{H}^j\textbf{F}|\Psi^{\sigma_0}_0}\nonumber\\
    &=\braket{\Psi^{\sigma_0}_0|[\textbf{F},\textbf{H}^i][\textbf{H}^j,\textbf{F}]|\Psi^{\sigma_0}_0}\,,
\end{align}
with $i+j=k$ and where the property $\textbf{H}\ket{\Psi^{\sigma_0}_0}=0$ has been used. Since $[\textbf{H}^n,\textbf{F}]=[H^n,F]$ for $n\in\mathbb{N}$, the bold notation can in fact be omitted. The needed commutators can be rewritten \cite{Gallatin83a} as
\begin{subequations}\label{eq:comms}
    \begin{align}
        [H^l,F]&=\sum_{n=0}^{l-1}\binom{l}{n}C_{l-n}H^n\,,\\
        [F,H^l]&=\sum_{n=0}^{l-1}\binom{l}{n}H^n\tilde{C}_{l-n}\,,
    \end{align}
\end{subequations}
with $C_l$ introduced in Eq.~\eqref{eq:def_comm} and $\tilde{C}_l$ defined through
\begin{align}\label{eq:def_comm_til}
\tilde{C}_l&\equiv\{{F},{H}^l\}\equiv[[...[[{F},\underbrace{{H}],{H}]...,{H}],{H}}_{\text{$l$ times}}]\,,
\end{align}
the two being equal up to a sign, i.e. $\tilde{C}_l=(-1)^lC_l$. The only non-vanishing contributions to Eq. \eqref{eq:dim_pass} are obtained for $n=0$ in Eqs. \eqref{eq:comms} by virtue of $\textbf{H}\ket{\Psi^{\sigma_0}_0}=0$ (even though the bold notation could be omitted in the meantime). This finishes to prove Eq. \eqref{eq:mk_gen}.

\section{Second-quantized operators}
\label{sec:op}

Given an arbitrary orthonormal basis of the one-body Hilbert space $\mathcal{H}_1$ represented by the particle annihilation and creation operators $\{c_{p},c^{\dagger}_{p}\}$, a generic (particle-number conserving) operator $O$ containing up to three-body operators reads as
\begin{align}
O &\equiv O^{[0]} + O^{[2]} + O^{[4]} + O^{[6]} \nonumber \\
&\equiv O^{00} + O^{11} + O^{22} + O^{33} \nonumber \\
&\equiv O^{00}  \nonumber \\
&+ \frac{1}{(1!)^2} \sum _{pq} o^{11}_{pq} c^{\dagger}_{p} c_{q}  \nonumber \\
&+ \frac{1}{(2!)^2} \sum _{pqrs} o^{22}_{pqrs}  c^{\dagger}_{p} c^{\dagger}_{q} c_{s} c_{r}   \nonumber \\
&+ \frac{1}{(3!)^2} \sum_{pqrstu} o^{33}_{pqrstu} c^{\dagger}_{p}c^{\dagger}_{q}c^{\dagger}_{r}c_{u}c_{t}c_{s} \, , \label{operatorO}
\end{align}
where $O^{[0]}=O^{00}$ is a number. Given that $O$ is presently taken to be particle-number conserving, the $k$-body class $O^{[2k]}$ contains a single operator $O^{kk}$ characterized by the equal number $k$ of particle-creation and annihilation operators.  Such an operator is obviously in normal order with respect to the particle vacuum.

Matrix elements entering Eq.~\eqref{operatorO} are fully antisymmetric, i.e.
\begin{equation}
o^{kk}_{p_1 \ldots p_{k} p_{k+1} \ldots p_{2k}} = (-1)^{\sigma(P)} o^{kk}_{P(p_1 \ldots p_{k} | p_{k+1} \ldots p_{2k})} 
\end{equation}
where $\sigma(P)$ refers to the signature of the permutation $P$. The notation $P(\ldots | \ldots)$ denotes a separation into the $k$ particle-creation operators and the $k$ particle-annihilation operators such that permutations are only considered between members of the same group.

\section{Operator $M_1(1,0)$}
\label{sec:mom_expl}

The algebraic expressions of the matrix elements defining the operator $M_1(1,0)$ allowing to $m_1$ via the GSEV approach are presently derived. All notations are consistent with \ref{sec:op} for operators expressed in normal order with respect to the particle vacuum.



There are two equivalent ways to obtain the odd-moment operators, namely given by Eqs. \eqref{eq:odd_operator} and \eqref{eq:sym_op}, respectively. They are explored separately below. 

\subsection{Similarity-transformed H}
\label{sec:sym_H}

Using Eq. \eqref{eq:sym_op} for $k=1$, the operator is given by 
\begin{equation}
    M_1(1,0)=-\frac{1}{2}\frac{\partial^2}{\partial\eta^2}H_1(\eta)\Bigg|_{\eta=0}
\end{equation}
with
\begin{equation}
    H_1(\eta)=e^{-\eta F}He^{\eta F}\,,
\end{equation}
such that
\begin{align}\label{eq:sym_inter}
    H_1(\eta)=&H_1^{[0]}(\eta)
    +\sum_{ab}h^{11}_{ab}\,e^{-\eta F}c^\dagger_ac_be^{\eta F}\nonumber\\
    +&\frac{1}{(2!)^2}\sum_{abcd}h^{22}_{abcd}\,e^{-\eta F}c^\dagger_ac^\dagger_bc_dc_ce^{\eta F}\nonumber\\
    +&\frac{1}{(3!)^2}\sum_{abcdef}h^{33}_{abcdef}\,e^{-\eta F}c^\dagger_ac^\dagger_bc^\dagger_cc_fc_ec_de^{\eta F}\,.
\end{align}
As shown below, the similarity transformation, $F$ being a one-body operator, does not change the rank of the operator, such that $M_1(1,0)$ has the same rank as $H$.   
Introducing the identity operator in-between each pair of creation and/or annihilation operators under the form 
\begin{equation}
    1=e^{\eta F}e^{-\eta F}\,,
\end{equation}
the similarity transformation is separately performed on each creation (annihilation) operator. The elementary commutator 
\begin{align}
    [F,c^\dagger_a]&=\sum_{kl}f^{11}_{kl}c_k^\dagger c_lc_a^\dagger-c_a^\dagger F\nonumber\\
    &=\sum_{kl}f^{11}_{kl}c_k^\dagger(\delta_{la}-c_a^\dagger c_l)-c_a^\dagger F\nonumber\\
    &=\sum_{k}f^{11}_{ka}c^\dagger_k\,,
\end{align}
together with Baker-Campbell-Hausdorff's formula allows one to obtain
\begin{subequations}
    \begin{align}
        e^{-\eta F}c_a^\dagger e^{\eta F}&=c_a^\dagger + \sum_{n=1}^\infty\frac{(-\eta)^n}{n!}[\underbrace{{F},[{F},...[{F},[{F}}_{\text{$n$ times}},{c_a^\dagger}]]...]]\nonumber\\
        &=c_a^\dagger-\eta\sum_{k}f^{11}_{ka}c^\dagger_k+\eta^2\frac{1}{2!}\sum_{kl}f^{11}_{lk}f^{11}_{ka}c^\dagger_l+\dots\nonumber\\
        &=c_a^\dagger-\eta\sum_{k}f^{11}_{ka}c^\dagger_k+\eta^2\frac{1}{2!}\sum_{k}(f^{11})^2_{ka}c^\dagger_k+\dots\nonumber\\
        &=\sum_k(e^{-\eta f^{11}})_{ka}c_k^\dagger\,.\\
        \intertext{Similarly, one has}
        e^{-\eta F}c_a e^{\eta F}&=\sum_k(e^{\eta f^{11}})_{ka}c_k\,.
    \end{align}
\end{subequations}
Eventually, Eq. \eqref{eq:sym_inter} is written as
\begin{align}
    H_1(\eta)=&H_1^{[0]}(\eta)\nonumber\\
    +&\sum_{kl}h^{11}_{kl}(\eta)\,c_k^\dagger c_l\nonumber\\
    +&\frac{1}{(2!)^2}\sum_{klmn}h^{22}_{klmn}(\eta)\,c^\dagger_kc^\dagger_lc_nc_m\nonumber\\
    +&\frac{1}{(3!)^2}\sum_{\substack{klm\\nop}}h^{33}_{klmnop}(\eta)\,c^\dagger_kc^\dagger_lc^\dagger_mc_pc_oc_n\,,
\end{align}
with the similarity-transformed matrix elements being defined as
\begin{align}
\label{eq:h_trans}
    h^{nn}_{k_1\dots k_n l_1\dots l_n}(\eta)&\equiv\sum_{\substack{a_1\dots a_n\\b_1\dots b_n}}h^{nn}_{a_1\dots a_n b_1\dots b_n}\nonumber\\
    &\times(e^{-\eta f^{11}})_{k_1 a_1}\dots(e^{-\eta f^{11}})_{k_n a_n}\nonumber\\
    &\times(e^{\eta f^{11}})_{l_1 b_1}\dots(e^{\eta f^{11}})_{l_n b_n}\,
\end{align}
and
\begin{equation}
    H_1^{[0]}(\eta)\equiv H^{[0]}\,.
\end{equation}
The similarity-transformed matrix elements on the left-hand side of Eqs. \eqref{eq:h_trans} naturally inherit the antisymmetry of the original matrix elements, as it can be checked directly. The second derivative with respect to $\eta$ can now be explicitly performed to derive the matrix elements of $M_1(1,0)$, i.e.
\begin{subequations}\label{eq:mat_el_sym}
    \begin{align}
    M^{[0]}_1=&0\,,\\
    m^{11}_{1,kl}(1,0)\equiv&-\frac{1}{2}\frac{\partial^2}{\partial\eta^2}h^{11}_{kl}(\eta)\Big|_{\eta=0}\,,\\
    m^{22}_{1,klmn}(1,0)\equiv&-\frac{1}{2}\frac{\partial^2}{\partial\eta^2}h^{22}_{klmn}(\eta)\Big|_{\eta=0}\,,\\
    m^{33}_{1,klmnop}(1,0)\equiv&-\frac{1}{2}\frac{\partial^2}{\partial\eta^2}h^{33}_{klmnop}(\eta)\Big|_{\eta=0}\,.
    \end{align}
\end{subequations}
As for the matrix elements of the similarity-transformed Hamiltonian in Eq.~\eqref{eq:h_trans}, the matrix elements of $M_1(1,0)$ are manifestly antisymmetric. The explicit writing of Eqs.~\eqref{eq:mat_el_sym} can be found in Sec.~4.3.2 of Ref.~\cite{Porro_thesis}. The results are strictly equivalent to those obtained via the commutator-based formulation described below.

\subsection{Commutator approach}
\label{sec:comm_m1}

The matrix elements of $M_1$ can also be derived from Eq.~\eqref{eq:m_operator} for $k=1$ (e.g. $i=0$ and $j=1$). This is achieved by applying Wick's theorem with respect to the particle vacuum $\ket{0}$. In this case the only non-vanishing contraction at play is
\begin{equation}
\contraction{}{c_a}{}{c_b}
    c_a\,\,c_b^\dagger\equiv\braket{0|c_ac_b^\dagger|0}=\delta_{ab}\,.
\end{equation}	
The commutator $C_1=[H,F]$ is computed separately for the various components of $H$. The operator $F$ being a one-body operator, the commutator preserves the $n$-body nature of the component $H^{[n]}$ such that each $n$-body component of $C_1$ is introduced as
\begin{subequations}\label{eq:o_n}	
\begin{align}
    [H^{[0]},F]=&0\,,\label{eq:c0}\\
    [H^{[1]},F]
    \equiv&\sum_{ab}c^{11}_{1,ab}\,c^\dagger_ac_b\,,\label{eq:comm_1b}\\
    [H^{[2]},F]
    \equiv&\frac{1}{(2!)^2}\sum_{abcd}c^{22}_{1,abcd}\,c^\dagger_ac^\dagger_bc_dc_c\,,\label{eq:comm_2b}\\
    [H^{[3]},F]
    \equiv&\frac{1}{(3!)^2}\sum_{\substack{abc\\def}}c^{33}_{1,abcdef}c^\dagger_ac^\dagger_bc^\dagger_cc_fc_ec_d\, .
\end{align}
\end{subequations}
The derivation of the matrix elements from Eqs.~\eqref{eq:o_n} relies on the tool developed in Ref.~\cite{tichai2021adg}. This tool allows one to compute the antisymmetrized matrix elements of the normal-ordered operator obtained via the commutator of any two normal-ordered operators. While the development was originally done with respect to a Bogoliubov vacuum $\ket{\Phi_{\!_\text{HFB}}}$ and expressing normal-ordered operators in the associated quai-particle basis, it can be readily exploited here by simply substituting quasi-particle operators $\beta^{\dagger}$ ($\beta$) with particle operators $c^\dagger$ ($c$) and by using the particle vacuum $\ket{0}$ instead of the Bogoliubov one. Naturally the particle formalism only needs to retain particle-number-conserving components. 

Eventually, the matrix elements of the elementary commutator from Eqs.~\eqref{eq:o_n} can be expressed as
\begin{subequations}\label{eq:mat_el_adg}
    \begin{align}
        c^{11}_{1,ab}&=\sum_{k}h^{11}_{ak}f^{11}_{kb}-\sum_{k}f^{11}_{ak}h^{11}_{kb}\,,\label{eq:c1_1b}\\
        c^{22}_{1,abcd}&=P(c/d)\sum_{{k}}h^{22}_{abc{k}}f^{11}_{{k}d}\nonumber\\
        &-P(a/b)\sum_{{k}}f^{11}_{a{k}}h^{22}_{{k}bcd}\,,\\
        c^{33}_{1,abcdef}&=P(de/f)\sum_{{k}}h^{33}_{abcde{k}}f^{11}_{{k}f}\nonumber\\
        &-P(a/bc)\sum_{{k}}f^{11}_{a{k}}h^{33}_{{k}bcdef}\,,
    \end{align}
\end{subequations}
with
\begin{subequations}
    \begin{align}
        P(a/b)&\equiv1-P_{ab}\,,\\
        P(a/bc)&\equiv1-P_{ab}-P_{ac}\,,\\
        P(ab/c)&\equiv1-P_{ac}-P_{bc}\,,
    \end{align}
\end{subequations}
and where $P_{ab}$ denotes the transposition operator exchanging indices $a$ and $b$. The extended writing of Eqs.~\eqref{eq:mat_el_adg}, i.e. with permutations explicitly carried out, can be found in Sec.~4.3.2 of Ref.~\cite{Porro_thesis}.

The above result is exploited to readily compute the nested commutator needed to obtain the $m_1$ operator
\begin{equation}\label{eq:M_1}
    M_1(1,0)\equiv-\frac{1}{2}[[H,F],F]=-\frac{1}{2}[C_1,F]\,,
\end{equation}
by substituting $H$ with $C_1$ in Eqs. \eqref{eq:mat_el_adg}. Eventually, the matrix elements of $M_1(1,0)$ are obtained as
\begin{subequations}
\label{eq:m1_contr}
    \begin{align}
    M_1^{[0]}&=0\,,\\
    m^{11}_{1,ab}&\equiv-\frac{1}{2}\sum_{c}\{c^{11}_{1,ac}f^{11}_{cb}-f^{11}_{ac}c^{11}_{1,cb}\}\,,\\
    m^{22}_{1,abcd}&=\frac{1}{2}P(a/b)\sum_{{k}}f^{11}_{a{k}}c^{22}_{{k}bcd}\nonumber\\&-\frac{1}{2}P(c/d)\sum_{{k}}c^{22}_{abc{k}}f^{11}_{{k}d}\,,\\
    m^{33}_{1,abcdef}&=\frac{1}{2}P(a/bc)\sum_{{k}}f^{11}_{a{k}}c^{33}_{{k}bcdef}\nonumber\\&-\frac{1}{2}P(de/f)\sum_{{k}}c^{33}_{abcde{k}}f^{11}_{{k}f}\,.
    \end{align}
\end{subequations}
The extended writing of Eqs.~\eqref{eq:m1_contr} is provided in Sec.~4.3.2 of Ref.~\cite{Porro_thesis} and is found to be identical to the similarity-evolved derivation from \ref{sec:sym_H}. 

\section{Strength function extraction}
\label{sec:relation_obs}

The actual relation of the strength function to scattering observables is hereby briefly discussed. At first order in perturbation theory, the transition rate $w_{0\rightarrow\nu}$ from the ground state $\ket{\Psi^{\sigma_0}_0}$ to an excited state $\ket{\Psi^{\sigma}_\nu}$ mediated by the time-independent operator $F$ is provided by Fermi's golden rule
\begin{equation}
w_{0\rightarrow\nu}=2\pi|\braket{\Psi^{\sigma}_\nu|F|\Psi^{\sigma_0}_0}|^2\delta(E^{\sigma}_\nu-E^{\sigma_0}_0-E)\, .
\end{equation}
The corresponding cross section $\sigma_{0\to\nu}$ is obtained normalising the transition rate by the flux of incident particles and the number of scattering centers
\begin{equation}
    \frac{d\sigma_{0\to\nu}}{dE}=w_{0\rightarrow\nu}\times\frac{1}{\text{flux}}\times\frac{1}{\text{\# of sc. centers}}\,.
\end{equation}
The total cross section is computed by summing over all possible final states $\nu$ so that it can be expressed as 
\begin{equation}
    \sigma=2\pi\int_{-\infty}^{+\infty}S(E)dE=2\pi m_0\,.
\end{equation}
In practice, double-differential cross sections are experimentally measured to perform a \textit{multipole-decomposition analysis} (MDA), allowing the extraction of the multipole strength distributions \cite{Garg2018a}. In the MDA process, the experimental cross-sections at each angle are binned into small (typically, $\leq 1$ MeV) excitation energy intervals. The laboratory angular distributions for each excitation-energy bin are then converted into the centre-of-mass frame using standard Jacobian and relativistic kinematics. For each excitation energy bin, the experimental angular distributions are fitted by means of the least-square method with the linear combination of the calculated double-differential cross sections associated to different multipoles:
\begin{equation}
    \frac{d^2\sigma^{\text{exp}}}{d\Omega dE}\bigg|_{E_x}=\sum_{L=0}^\infty a_L(E_x)\frac{d^2\sigma^{{\text{DWBA}}}_L}{d\Omega dE}\bigg|_{E_x}\,,
\end{equation}
where $a_L(E_x)$ is the $m_1$ sum rule fraction for the $L$-th component. The cross sections used for the fit procedure correspond to the 100\% of $m_1$ for the $L$-th multipole at excitation energy $E_x$ calculated using the distorted-wave Born approximation (DWBA). In such calculations an optical potential is used as the scattering potential. The fractions of $m_1$, $a_L(E_x)$, for various multipole components are determined by minimising $\chi^2$ error. Eventually, the strength distributions for different multipolarities are obtained by multiplying the extracted $a_L(E_x)$'s by the strength corresponding to 100\% $m_1$ at the given energy $E_x$ 
\begin{equation}
    S_L(E_x)=\frac{m_{L,1}}{E_x}a_L(E_x)\,.
\end{equation}

Traditionally, the energy-weighted sum rules $m_{L,1}$ employed in the above procedure for different $L$'s are always the textbook $\text{EWSR}_{\text{lab}}$ rather than the appropriate intrinsic one discussed in Sec.~\ref{sec:ewsr_ex}.

\section{Intrinsic EWSR}
\label{correctCMEWSR}

The monopole EWSR from Eq.~\eqref{eq:EWSR0} is evaluated under the assumption that only the kinetic energy $T_{\text{lab}}$ from Eq.~\eqref{eq:t_lab} contributes  to Eq.~\eqref{eq:odd_operator}, such that
\begin{align}
    \text{EWSR}_{\text{lab}}(r^2)&=-\frac{1}{2}\braket{[[T_{\text{lab}},r^2],r^2]}_{\text{gs}} \nonumber \\
    &= \frac{2\hbar^2A}{m}\braket{r^2_{\text{lab}}}_{\text{gs}} \,.
\end{align}
The correction to $\text{EWSR}_{\text{lab}}(r^2)$ due to the subtraction of the center-of-mass kinetic energy $T_{\text{cm}}$ (Eq.~\eqref{eq:t_cm})  from the Hamiltonian is given by
\begin{align}
\label{eq:ewsr_dim}
    \delta m_1^{\text{cm}}(r^2)&\equiv -\frac{1}{2}\braket{[[T_{\text{cm}},r^2],r^2]}_{\text{gs}}\nonumber \\
    &=\frac{1}{4mA}\sum_{ijkl=1}^A\braket{[[\vec{p}_i\cdot\vec{p}_j,r_k^2],r_l^2]}_{\text{gs}}\,.
\end{align}
The commutator relation
\begin{equation}
    [\vec{p}_i,f(\vec{r}_k)]=-i\hbar\vec{\nabla}_if(\vec{r}_k)
\end{equation}
is employed to evaluate the elementary commutator
\begin{equation}
    [\vec{p}_i,r^2_k]=-2i\hbar\vec{r}_i\delta_{ik}\, ,
\end{equation}
allowing to process Eq.~\eqref{eq:ewsr_dim}  according to
\begin{align}
    \delta m_1^{\text{cm}}(r^2)&=\frac{1}{4mA}\sum_{ijkl}\braket{[\vec{p}_i\cdot[\vec{p}_j,r_k^2],r_l^2]+[[\vec{p}_i,r_k^2]\cdot\vec{p}_j,r_l^2]}_{\text{gs}}\nonumber\\
    &=\frac{1}{4mA}\sum_{ijkl}\braket{[\vec{p}_i,r_l^2]\cdot[\vec{p}_j,r_k^2]+[\vec{p}_i,r_k^2]\cdot[\vec{p}_j,r_l^2]}_{\text{gs}}\nonumber\\
    &=-\frac{2\hbar^2}{mA}\sum_{ij}\braket{\vec{r}_i\cdot\vec{r}_j}_{\text{gs}}\nonumber\\
    &=-\frac{2\hbar^2 A}{m} \braket{R^2_{\text{cm}}}_{\text{gs}} \,,
\end{align}
where the second equality follows from 
\begin{align}
    [[\vec{p}_i,r^2_k],r^2_l]&=-2i\hbar[\vec{r}_i,r^2_l]\delta_{ik}=0\, ,
\end{align}
and where
\begin{align}
    \vec{R}_{\text{cm}} &\equiv \frac{1}{A} \sum_{i=1}^{A} \vec{r}_i \, ,
\end{align}
denotes the center of mass coordinate.

Eventually, the EWSR associated with the intrinsic Hamiltonian  can be written under the alternating forms
\begin{align}
    \text{EWSR}_{\text{int}}(r^2)&=\text{EWSR}_{\text{lab}}(r^2)+\delta m_1^{\text{cm}}(r^2)\nonumber\\
    &= \frac{2\hbar^2A}{m} \braket{r^2_{\text{lab}}-R^2_{\text{cm}}}_{\text{gs}} \nonumber \\
    &= \frac{2\hbar^2A}{m} \braket{r^2_{\text{int}}}_{\text{gs}} \nonumber \\
    &=\text{EWSR}_{\text{lab}}(r^2)\Big(1-\frac{1}{A}\Big) \nonumber \\
    &\,\,\,\, -\frac{2\hbar^2}{mA}\sum_{i\neq j}\braket{\vec{r}_i\cdot\vec{r}_j}_{\text{gs}} \, . \label{finalEWSRint}
\end{align}
The above result demonstrates that the subtraction of $T_{\text{cm}}$ in $H$ leads to replacing the laboratory-frame mean-square radius  $\braket{r^2_{\text{lab}}}$ by the intrinsic one $\braket{r^2_{\text{int}}}$. The last line splits $\delta m_1^{\text{cm}}(r^2)$ into its one- and a two-body contributions to demonstrate that the one-body part of $T_{\text{cm}}$ leads to a simple A-dependent renormalization of $\text{EWSR}_{\text{lab}}(r^2)$~\cite{colo2013CPC}. 

\bibliography{biblio.bib}

\end{document}